\documentclass[final,3p,times]{elsarticle}
\usepackage[english]{babel}
\usepackage{graphics}
\usepackage[utf8x]{inputenc}
\usepackage{epsfig}
\usepackage{subfigure}
\usepackage{lscape}
\usepackage{amssymb}
\usepackage{amsmath}
\usepackage{tabularx}
\usepackage{pdflscape}
\usepackage{subeqnarray}
\usepackage{color}

\usepackage{epstopdf}
\usepackage{xcolor}

\usepackage{graphicx}
\usepackage{caption}
\usepackage{multirow}

\begin{document}

\begin{frontmatter}


\author{F. Municchi$^1$}
\author{I. El Mellas$^2$}
\author{O. K. Matar$^3$}
 \author{M. Magnini\corref{*}$^2$}

 \cortext[*]{Corresponding author. E-mail: mirco.magnini@nottingham.ac.uk}
\address{$^1$ Colorado School  of Mines, 1500 Illinois St., Golden, CO 80401 \\ \vspace{0.2cm} $^2$ Department of Mechanical, Materials and Manufacturing Engineering, University of Nottingham, Nottingham NG7 2RD, United Kingdom \\ \vspace{0.2cm} $^3$ Department of Chemical Engineering, Imperial College London, London SW7 2AZ, United Kingdom}

\title{Conjugate heat transfer effects on flow boiling in microchannels}

\begin{abstract}

This article presents a computational study of saturated flow boiling in non-circular microchannels. The unit channel of a multi-microchannel evaporator, consisting of the fluidic channel and surrounding evaporator walls, is emulated and the conjugate heat transfer problem is solved. Simulations are performed using OpenFOAM v2106 and the built-in geometric Volume Of Fluid method, augmented with self-developed libraries to include liquid-vapour phase-change and improve the surface tension force calculation. 
A systematic study is conducted by employing water at atmospheric pressure, a channel hydraulic diameter of $D_h=229\,\mathrm{\mu m}$, a uniform base heat flux of $q_b=100\,\mathrm{kW/m^2}$, and by varying the channel width-to-height aspect-ratio and channel fin thickness in the range $\epsilon=0.25-4$ and $W_f=D_h/8-D_h$, respectively. The effects of conjugate heat transfer and channel aspect-ratio on the bubble and evaporative film dynamics, heat transfer, and evaporator temperature are investigated in detail. 
This study reveals that, when the flow is single-phase, higher Nusselt numbers and lower evaporator  temperatures are achieved for $\epsilon <1$. 
In the two-phase flow regime, the trends of the Nusselt number versus the aspect-ratio are mixed, although for smaller channel fins an ascending trend of $\mathrm{Nu}$ for increasing aspect-ratios is apparent. Nonetheless, due to conjugate heat transfer, Nusselt numbers and evaporator base temperatures follow different trends when varying the aspect-ratio, and channels with $\epsilon<1$ seem to promote lower evaporator temperatures than higher aspect-ratio conduits, despite exhibiting slightly worse two-phase convective heat transfer performances.
\end{abstract}


\end{frontmatter}

%
%




\section{Introduction}
\label{Sec:intro}

Flow boiling in mini- and microchannels is recognised as one of the most efficient cooling solutions for high-power-density applications. Recent advances in manufacturing technology have enabled devices such as high-performance computers, power electronics, lasers, avionics, electric vehicles, batteries, photovoltaics, miniature fuel cells, energy conversion and storage systems, evaporators, condensers and reactors, amongst other, to operate at high power densities \cite{karayiannis2017}. These applications involve heat fluxes of $O (\mathrm{MW/m^2})$, while the heat removal capability of traditional single-phase cooling is below $1\,\mathrm{MW/m^2}$ \cite{tullius2011}, which has resulted in a dramatic and urgent demand for high-performance thermal management systems that can transfer unprecedentedly high heat fluxes. Boiling two-phase flows in microchannels \cite{agostini2007}: (i) yield very high heat transfer coefficients and maintain uniform surface temperatures, vital for the correct operation of components; (ii) respond passively to alleviate localised ‘hot-spots’, as the heat transfer coefficient increases with the heat flux without the need for actively-controlled (higher) flow rates; (iii) offer a large surface-to-volume ratio, benefitting the compactness of the system. As such, flow boiling in microchannels has been studied extensively in recent years, with a focus on flow pattern transitions, void fraction, pressure drop, heat transfer coefficient, and critical heat flux \cite{karayiannis2017,cheng2017}.

In order to increase the surface area, the two-phase flow is usually organised into microevaporators, where multiple parallel microchannels are manufactured into a thin metal die made of conductive material, which is placed in direct contact with the surface to be refrigerated \cite{alZaidi2021}.  Heat is delivered to the fluid via heat conduction through the solid walls of the evaporator and heat convection at the contact surface between fluid and solid, thus forming a conjugate heat transfer problem. Since heat is applied to one side of the evaporator, the channels are subject to a nonuniform heating condition, as heat is unequally distributed among the channel walls. The heat removal capability of a microevaporator is characterised by the convective heat transfer achieved by the two-phase flow, quantified by a boiling heat transfer coefficient or Nusselt number, and by the value of the evaporator base temperature, since the heat sink must limit the temperature of the device to be refrigerated below a threshold value. 

Within microchannels, after nucleation vapour bubbles grow quickly and occupy the cross-section  leading to slug or annular flow patterns, with bubbly flows being suppressed already at very low values of vapour quality \cite{harirchian2009}. Surface tension forces rearrange the liquid-vapour interface pattern into thick liquid lobes at channel corners and thin liquid films at the centre \cite{wong1995a}, and the thickness and morphology of this film strongly depends on the channel aspect-ratio \cite{deLozar2008,magnini2022a}. The distribution of this liquid film has a direct impact on the heat transfer coefficient, as both experimental \cite{han2012,rao2015} and numerical \cite{ferrari2018,magnini2020a} studies have demonstrated that the local heat transfer coefficient $h$ is inversely proportional to the film thickness $\delta$, $h \approx \lambda_l/\delta$, with $\lambda_l$ being the liquid thermal conductivity, with film dryout being highly detrimental to heat transfer performance. As such, the channel aspect-ratio is expected to yield significant impact on microchannel boiling heat transfer. The experimental literature on the impact of the channel shape on boiling heat transfer is rather vast, and comprehensive reviews have been carried out by \citet{magnini2020a,vontas2021} and \citet{alZaidi2021}. It emerges that there is still substantial disagreement on the effect of the channel aspect-ratio on boiling heat transfer, with  contrasting trends of heat transfer coefficient and microchannel wall temperature versus aspect-ratio being reported in these studies. 

More recently, interface-resolving numerical methods have been employed to investigate relevant fluid mechanics structures and heat transfer mechanisms pertinent to flow boiling in noncircular microchannels. \citet{magnini2020a} performed a systematic analysis of the impact of the channel aspect-ratio on the bubble dynamics and heat transfer and concluded that square channels performed better at lower flow rates while rectangular channels exhibited larger Nusselt numbers at higher flow rates. However, their study did not include conjugate heat transfer through the evaporator walls, and it considered an idealised slug flow at low heat flux conditions ($q \sim 10\,\mathrm{kW/m^2}$), where liquid film dryout never occurred. \citet{vontas2021} simulated flow boiling in single and multiple rectangular channels for different channel hydraulic diameters also accounting for the evaporator walls. They reported an ascending trend of the heat transfer coefficient when reducing the channel size, but did not study the impact of the channel aspect-ratio, which was maintained constant. \citet{lin2021} modeled flow boiling in a single rectangular microchannel accounting for one evaporator wall. They varied the material and thickness of the wall and observed that two-phase heat transfer was enhanced by thicker walls, which exhibited higher temperature and increased the bubble growth rate, and by highly-conductive materials. However, their study considered only one microchannel wall and thus it is not representative of an actual heat sink.

The literature review outlined above emphasises the fact that the impact of the channel aspect-ratio and conjugate heat transfer on flow boiling in microchannel evaporators is still unclear. Conjugate heat transfer is particularly important in microgeometries, where solid walls are of thickness comparable to the channel size, such that heat diffuses along all coordinate directions \cite{szczukiewicz2014} and the fluid is nonuniformly heated around the channel perimeter.  This article presents a computational study of the effect of the geometrical features of channel and evaporator walls on heat transfer in both single-phase and two-phase flow. The unit channel of a multi-microchannel evaporator, consisting of channel and surrounding walls, is modelled and a conjugate heat transfer problem is solved. Simulations are performed with OpenFOAM v2106, using the built-in geometric Volume Of Fluid (VOF) solver isoAdvector \cite{roenby2016}, augmented with self-developed functions implementing thermally-driven liquid-vapour phase-change and improving the native  surface tension method. Simulations are run for a constant value of the channel hydraulic diameter, $D_h=229\,\mathrm{\mu m}$, mass flux ($G=140\,\mathrm{kg/(m^2 s)}$) and base heat flux ($q_b=100\,\mathrm{kW/m^2}$), using water at atmospheric pressure and copper as working fluid and evaporator material, respectively. A constant heat flux is provided to the outer wall of the evaporator base, and the resulting single- and two-phase dynamics and heat transfer are investigated for a range of aspect-ratios $\epsilon=0.25-4$ and of channel fin widths $W_f=D_h/8-D_h$. 

The rest of this article is organised as follows: the numerical framework is described in Sec.~\ref{Sec:numModel} and results of validation benchmarks are illustrated in Sec.~\ref{Sec:validation}; the results of the single-phase and flow boiling simulations are presented in Sec.~\ref{Sec:results}; Sec.~\ref{Sec:disc} provides a discussion of the observed heat transfer trends, and conclusions are summarised in Sec.~\ref{Sec:concl}.

\section{Numerical framework}
\label{Sec:numModel}

\subsection{Governing equations}
\label{Sec:eqs}

The numerical model is based on the solution of a conjugate heat transfer problem in the fluid and solid regions of the domain. The fluid model solves the Navier-Stokes and energy equations for the flow of two immiscible phases, namely liquid and vapour, separated by an interface. The liquid and vapour phases are both treated as incompressible, Newtonian fluids. A single-fluid formulation is adopted and the two phases are treated as a single mixture fluid with variable properties across the interface, such that a single field of velocity, pressure, and temperature are sufficient to describe the flow, and a single set of conservation equations holds throughout the domain \cite{scardo_book}. Accordingly, the governing equations of mass, momentum, and energy, for a flow with phase-change, are expressed as follows:

\begin{equation}
\nabla \cdot \boldsymbol{u}=\frac{\dot{\rho}}{\rho}
\label{mass} 
\end{equation}
\begin{equation}
 \frac{\partial (\rho \boldsymbol{u})}{\partial t}+\nabla \cdot (\rho \boldsymbol{u}  \boldsymbol{u}) =-\nabla p+\nabla \cdot \mu \biggl[  ( \nabla \boldsymbol{u})+(\nabla \boldsymbol{u})^T  \biggr] + \boldsymbol{F_{\sigma}}   
 \label{momentum} 
\end{equation}
\begin{equation}
\frac{\partial (\rho c_pT)}{\partial t}+\nabla \cdot (\rho c_p\boldsymbol{u}T) = \nabla \cdot (\lambda \nabla T) + \dot{h}
\label{energy}
\end{equation}
where $\boldsymbol{u}$ indicates the fluid velocity, $\dot{\rho}$ the mass flux due to phase-change, $\rho$ the mixture fluid density, $t$ the time, $p$ the pressure, $\mu$ the dynamic viscosity, $\boldsymbol{F_{\sigma}}$ the surface tension force vector, $T$ the temperature, $c_p$ the constant pressure specific heat, $\lambda$ the thermal conductivity, and $\dot{h}$ is the enthalpy source due to phase-change. Details of surface tension and phase-change models are provided in the subsections below. Gravitational effects are neglected in this work.

By means of the VOF method, a volume fraction field $\alpha$ is defined to map liquid and vapour phases throughout the flow domain. In each computational cell of the domain, $\alpha$ identifies the fraction of the cell occupied by the primary phase, which corresponds to liquid in the present case. Therefore, the volume fraction takes values of 1 in the liquid, 0 in the vapour, and $0<\alpha<1$ in cells that are cut by the interface. The volume fraction field is evolved upon solution of the following transport equation:
\begin{equation}
\frac{\partial  \alpha}{\partial t}+\nabla \cdot ( \alpha \boldsymbol{u})=\frac{\dot{\rho}}{\rho}\alpha \label{alpha} 
\end{equation}
Using the volume fraction field, the properties of the mixture fluid can be computed as an average over the two phases, e.g. $\rho=\alpha \rho_l+(1-\alpha) \rho_v$, with the subscripts $v$ and $l$ denoting vapour- and liquid-specific properties, respectively. All the fluid-specific properties (e.g. $\rho_v$, $\rho_l$, $\mu_v$, $\mu_l$, etc.) are considered constant in this work.

A separate mesh is used to discretise the solid domain. The temperature field $T_s$ in the solid domain is obtained by solving the following heat conduction equation:
 \begin{equation}
\frac{\partial (\rho_s c_{p,s} T_s)}{\partial t}= \nabla \cdot (\lambda_s \nabla T_s)
\label{energys}
\end{equation}
where the subscript $s$ refers to solid. At the boundary between solid and fluid regions, the fluid and solid temperature fields are coupled by imposing continuity of temperature and heat flux, which is achieved by solving the fluid and solid energy equations iteratively, and adjusting the boundary conditions after every iteration.

\subsection{Surface tension model}
\label{Sec:stModel}

The surface tension force, $\boldsymbol{F_{\sigma}}$ in Eq.~\eqref{momentum}, is formulated according to the Continuum Surface Force method \cite{brackbill} and computed as:
\begin{equation}
\boldsymbol{F_{\sigma}}=\frac{2\rho}{\rho_l+\rho_v} \sigma \kappa |\nabla \alpha|
\label{surfTens} 
\end{equation}
where $\sigma$ is the surface tension coefficient (considered constant) and $\kappa$ the local interface curvature; the term $2\rho/(\rho_l+\rho_v)$ represents a density-correction factor that does not change the integral of the surface tension across the interface, but redistributes the surface tension towards the denser fluid to prevent unphysical accelerations in the region occupied by the
lighter fluid \cite{brackbill}; the impact of the density-correction term on the simulation of flow boiling in microchannels will be investigated in Sec.~\ref{Sec:muk}. The interface curvature  is estimated by means of derivatives of a smoothed volume fraction field $\tilde{\alpha}$, $\kappa=\nabla \cdot (\nabla \tilde{\alpha}/|\nabla \tilde{\alpha}|)$ \cite{hoang2013}, where the smoothed volume fraction field is obtained by interpolating $\alpha$ from the computational cell centres to face centres and by averaging the resulting $\alpha_f$ field back to cell centres according to: 
\begin{equation}
\tilde{\alpha}=\frac{\sum_{f} \alpha_f S_f}{\sum_{f}  S_f}
\label{surfTens} 
\end{equation}
where the sum spans the $f$ faces (each of area $S_f$) of the control volume. The use of a smoothed volume fraction field to calculate $\kappa$ was already included in the original work of \citet{brackbill} and has the effect of improving the accuracy of the calculation of $\kappa$, compared to using the unsmoothed field $\alpha$, thus reducing the magnitude of the spurious velocity caused by errors in the surface tension calculation. The smoothing cycle can be repeated multiple times to further smoothen $\alpha$ before calculating $\kappa$, though its beneficial effect on mitigating the spurious velocity saturates after a few cycles \cite{hoang2013}; the optimal number of smoothing cycles will be object of the analysis in Sec.~\ref{Sec:muk}.

A static contact angle is imposed at the wall boundaries where the liquid-vapour interface comes in contact with solid walls. The contact angle is set by adjusting the direction of the unit normal to the interface in boundary cells that are cut by the contact line, using OpenFOAM's built-in implementation.

\subsection{Phase-change model}

The mass and enthalpy source terms due to evaporation, $\dot{\rho}$ in Eqs.~\eqref{mass} and \eqref{alpha} and $\dot{h}$ in Eq.~\eqref{energy}, are modelled according to the work of \citet{hardt}. The evaporating mass flux at the liquid-vapour interface, here denoted as $\dot{m}$, is calculated as a function of the local interface superheat according to the Hertz-Knudsen-Schrage relationship \cite{carey}, and adopting the linearisation proposed by \citet{tanasawa} for low values of the superheat:
\begin{equation}
 \dot{m}=\frac{2\gamma}{2-\gamma} \biggl( \frac{M}{2 \pi R_g} \biggr)^{1/2} \frac{\rho_v h_{lv} (T_{lv}-T_{sat})}{T_{sat}^{3/2}}
\label{kin_mob}
\end{equation}
where $\gamma$ is the evaporation coefficient, $h_{lv}$ is the vaporisation latent heat, $M$ is the molecular weight of the fluid, $R_g$ the universal gas constant, $T_{lv}$ is the temperature at the liquid-vapour interface, and $T_{sat}$ the saturation temperature of the fluid. In this work, the evaporation coefficient is set to 1 according to preliminary test benchmarks \cite{magnini2013a,ferrari2018}. An initial volumetric mass source $\dot{\rho}_0$ is estimated by accounting for the evaporating mass flux calculated based on the temperature on the liquid side of the interface:
\begin{equation}
 \dot{\rho}_0=N \alpha |\nabla \alpha| \dot{m}
 \label{dotrho0}
\end{equation}
where $\alpha$ is the liquid volume fraction and $N$ is a normalisation factor to ensure that the global evaporation rate is preserved \cite{hardt}. A smoothed evaporation mass source $\dot{\rho}_1$ is then obtained by solving a steady diffusion equation as described in detail in \cite{hardt,ferrari2018}; the smoothing of the evaporation source term occurs over a few cells across the interface and improves the numerical stability of the solver. The final volumetric source $\dot{\rho}$ is obtained by redistributing $\dot{\rho}_1$ across the interface on the vapor and liquid side according to:
\begin{equation}
\dot{\rho}=
\begin{cases} N_v(1-\alpha)\dot{\rho}_1, \quad & \text{if}\, \alpha<\alpha_{cut}  \\ 
-N_l \alpha \dot{\rho}_1, \quad & \text{if}\, \alpha>1-\alpha_{cut}    \\ 
0, \quad & \text{if}\,\alpha_{cut}< \alpha<1-\alpha_{cut}
\end{cases}
\end{equation}
where $N_l,N_v$ are normalisation factors ensuring that the masses of liquid evaporated and vapour created are conserved by the redistribution step. The threshold parameter $\alpha_{cut}$, here $\alpha_{cut}=10^{-3}$, guarantees that the evaporation source term is nonzero only on vapour- or liquid-full cells. The enthalpy source $\dot{h}$ accounts for the latent heat dissipated by the evaporation process and is calculated from the initial volumetric source term as $\dot{h}=-\dot{\rho}_0 h_{lv}$. 

\subsection{Discretisation methods}

The governing equations are solved using a customised implementation in OpenFOAM, version v2106. The transport equations are discretised with a finite-volume method on a collocated grid arrangement. OpenFOAM v2106 provides both an algebraic (interFOAM; see \cite{weller2008,deshpande2012}) and a geometric (isoAdvector; see \cite{roenby2016,scheufler2019}) VOF solver. The numerical framework outlined in the previous subsections has been implemented in both versions of VOF, which in our custom solver are merged into a single solver, where the interface advection method is specified by the user as a new solver option in fvSolution. This way, the user can select whether to run the solver using the algebraic VOF (interFOAM mode), or using the geometric VOF (isoAdvector mode) with any of the built-in interface reconstruction methods (isoAlpha, isoRDF, plicRDF; see \cite{scheufler2019}). When running the solver in interFOAM mode, the volume fraction Eq.~\eqref{alpha} is discretised with a first-order time-explicit scheme based on the MULES (Multidimensional Universal Limiter for Explicit Solution) algorithm \cite{weller2008,deshpande2012}, set with \textit{nAlphaCorr 1}, \textit{nAlphaSubCycles 3} and \textit{cAlpha 1}. When running the solver in isoAdvector mode, the advection method \textit{isoAlpha} is selected, with \textit{nAlphaSubCycles 3}. All other equations are integrated in time with a first-order implicit method. The divergence operators are discretised using second-order TVD (Total Variation Diminishing) schemes \cite{muscl}, whereas Laplacian operators are discretised with central finite-differences. The PISO (Pressure Implicit Splitting of Operators) algorithm \cite{piso} is utilised to iteratively update pressures and velocities (\textit{momentumPredictor yes}; \textit{nCorrectors 3}) within each time-step. The residuals thresholds for the iterative solution of the flow equations are set to $10^{-7}$ for the velocity, $10^{-9}$ for the pressure, $10^{-8}$ for the volume fraction, and $10^{-10}$ for the temperature and the evaporation source term smoothing equation. The time-step of the simulation is variable and is calculated based on a maximum allowed Courant number of $0.1$, unless otherwise specified.

\section{Validation}
\label{Sec:validation}

\subsection{Vapour bubble growth in superheated liquid}
\label{Sec:scriven}

To assess the implementation of the evaporation model in both the algebraic and geometric VOF methods, the first test case selected is the growth of a spherical vapour bubble in an infinitely extended superheated liquid domain. When a vapour bubble nucleates in a pool of superheated liquid ($T_l=T_\infty$) far from solid walls, there exists an initial growth stage where $T_v=T_\infty$ and $p_v=p_{sat}(T_\infty)$, the bubble growth is limited only by the inertia of the surrounding liquid being displaced and the bubble radius follows a linear trend with time. As the bubble grows, the vapour pressure and temperature decrease until $p_v \approx p_\infty$ and $T_v \approx T_{sat}(p_\infty)$. This identifies a second growth stage where the bubble growth rate is limited only by the rate at which latent heat is supplied at the interface, and the bubble radius obeys a $\sim t^{1/2}$ law \cite{carey}. 

\begin{figure}[t!]
\begin{center}
\includegraphics[width=160mm]{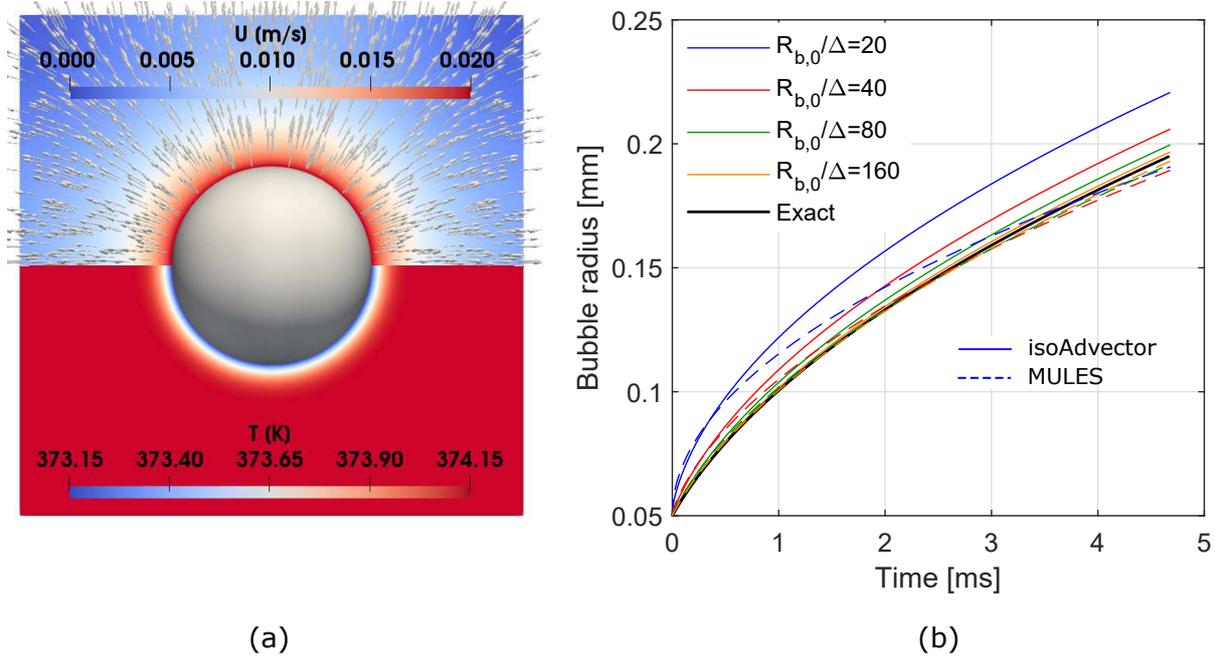}
\caption{Vapour bubble growth in superheated liquid. (a) Snapshot of the bubble (grey surface), velocity field (top half) and temperature field (bottom half) at the end of a simulation run using the geometric VOF and $R_{b,0}/\Delta=40$. (b) Vapour bubble radius versus time for simulations run with both the algebraic (MULES) and geometric (isoAdvector) VOF methods; $R_{b,0}$ indicates the bubble radius at $t=0$ and $\Delta$ the cell size.}
\label{scriven}
\end{center}
\end{figure}

The latter heat-transfer-controlled growth stage is reproduced here as a validation benchmark for the solver. A spherical steam bubble of initial radius $R_{b,0}=50\,\mathrm{\mu m}$ and temperature $T_v=373.15\,\mathrm{K}$ is placed in a pool of superheated water at the system pressure $p_\infty=1\,\mathrm{atm}$ and temperature $T_\infty=374.15\,\mathrm{K}$. The fluid domain is a two-dimensional axisymmetric square box of side of $1\,\mathrm{mm}$; using symmetry boundary conditions, only one quarter of the domain is simulated. A thin thermal boundary layer surrounds the bubble during its growth and the exact temperature profile within the liquid must be set at $t=0$ for a coherent initial growth. \citet{scriven} has derived an analytical solution for the heat-transfer-controlled growth stage of the bubble, providing expressions for the liquid temperature over time and distance from the interface, and a law for the bubble radius over time $R_b(t)=2\beta(at)^{1/2}$, with $\beta$ being a constant obtained from the solution \cite{scriven}, and $a=\lambda_l/(\rho_l c_{p,l})$ being the liquid thermal diffusivity. For the conditions presently simulated, the initial boundary layer thickness (when, at $t=0$, $R_{b,0}=50\,\mathrm{\mu m}$) is about $16\,\mathrm{\mu m}$ and the temperature profile resulting from the analytical solution is set as an initial condition for the liquid temperature in the simulation. The initial volume fraction field is set via OpenFOAM's utility setAlphaField, which enables accurate initialisation of volume fractions by calculating the intersections between implicit functions defining the bubble shape and the domain mesh. The simulations are run till $t=0.0047\,\mathrm{s}$, where the bubble is about 4 times its initial size. The maximum Courant number allowed for the simulation is set to 0.02.

The numerical solver is run both in algebraic and geometric VOF modes to compare the two different interface advection methods. Surface tension is disabled by setting $\sigma=0$, thus enabling us to test specifically the performances of the phase-change model and interface advection methods, without the influence of the surface tension algorithm. Four different meshes are utilised, all structured with orthogonal and uniform hexahedrons, with grid spacings of $\Delta=2.5\,\mathrm{\mu m}$ ($R_{b,0}/\Delta=20$), $\Delta=1.25\,\mathrm{\mu m}$ (40), $\Delta=0.625\,\mathrm{\mu m}$ (80), and $\Delta=0.3125\,\mathrm{\mu m}$ (160). A snapshot of the bubble growth dynamics is shown in Fig.~\ref{scriven}(a), where the velocity and temperature fields surrounding the bubble at the last time instant of the simulation are depicted. It can be seen that the bubble preserves the spherical shape during its growth, a thin temperature boundary layer surrounds the bubble and the liquid velocity is directed radially outward, with a maximum magnitude identified in the proximity of the liquid-vapour interface. The comparison of the bubble growth rate versus time for the different solver configurations tested and the analytical solution are presented in Fig.~\ref{scriven}(b). At low mesh resolutions, all advection methods yield a faster growth rate during the initial growth stage. This can be ascribed to the insufficient resolution of the initial thermal boundary layer surrounding the bubble. At $t=0$, the evaporation rate is calculated based on the temperatures at the centroids of the first few liquid cells nearby the interface, and therefore when the mesh is coarser these centroids are farther from the interface and experience excessively high temperatures, resulting in higher evaporation rates. At low mesh resolutions, the combination of phase-change model and algebraic VOF method seems to perform better than the geometric VOF in terms of $R(t)$, however the slopes of the $R(t)$ curve exhibited by the MULES method at later growth stages are more far off the exact solution than the isoAdvector ones. The deviations with the exact solution become of comparable magnitude between the two methods when $R_{b,0}/\Delta=80$ and isoAdvector (interFOAM) overpredicts (underpredicts) the exact growth rate by less than 2\%, whereas analytical and numerical solutions become almost indistinguishable when $R_{b,0}/\Delta=160$. Therefore, both methods yield solutions that converge to the analytical one as the mesh is refined.

 In terms of computational overhead, the simulations performed with the algebraic VOF were slightly faster than those run with isoAdvector, which can be ascribed to the extra geometric interface reconstruction steps performed by the latter. The simulations with $R_{b,0}/\Delta=80$ (160), featuring 640,000 (2,560,000) mesh cells, required 276 (3,100) core-hours for the algebraic VOF and 320 (3,950) core-hours for the geometric VOF. These simulations were run on UK's Tier-2 supercomputer Sulis, which features Dell PowerEdge R6525 computing nodes each with two AMD EPYC 7742 (Rome) 2.25 GHz 64-core processors, thus making 128 cores and 512 GB DDR4-3200 RAM per node.

\subsection{Flow boiling in a microchannel}
\label{Sec:muk}

\begin{figure}[b!]
\begin{center}
\includegraphics[width=160mm]{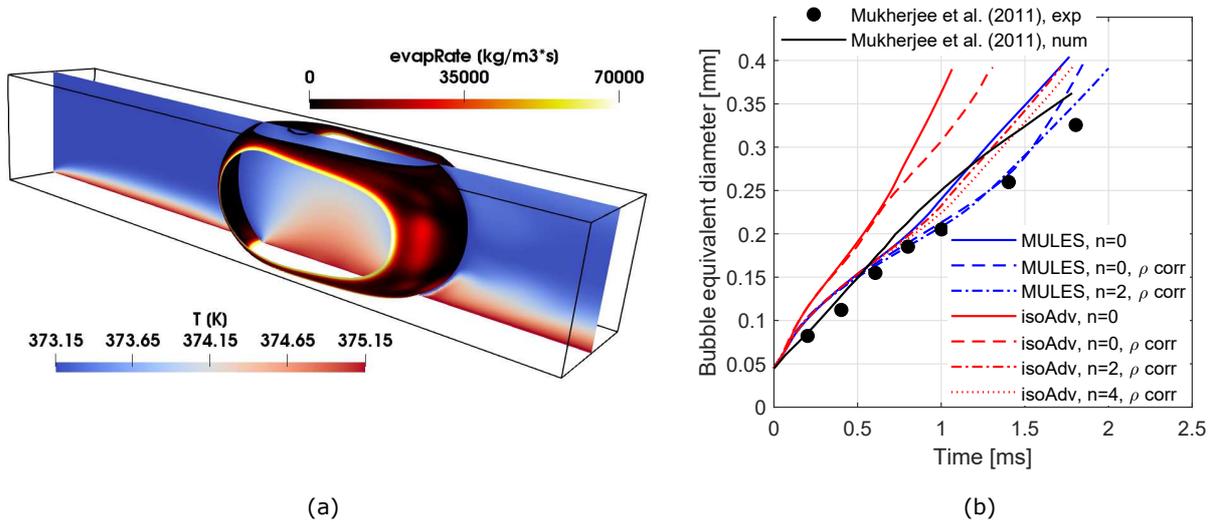}
\caption{Vapour bubble growth in a microchannel, comparison with \citet{mukherjee2011}. (a) Snapshot of the bubble, coloured by the evaporation rate, and temperature field along a vertical centreline plane, at $t=1.7\,\mathrm{ms}$ for a simulation run using the geometric VOF, $n=4$ curvature smoothers and density correction. Flow is from left to right. (b) Bubble equivalent diameter versus time for simulations run with both the algebraic (MULES) and geometric (isoAdv) VOF methods, different number of smoothers ($n$) in the curvature calculation, with or without density-correction term ($\rho$ corr) for the surface tension.}
\label{mukherjee}
\end{center}
\end{figure}

The second validation benchmark selected is the numerical simulation of the growth of a vapour bubble at the wall of a heated microchannel, which emulates the experiments and matching simulation performed by \citet{mukherjee2011}. \citet{mukherjee2011} performed an experiment in a microchannel of slightly trapezoidal cross-section and hydraulic diameter $D_h=229\,\mathrm{\mu m}$, where saturated water at atmospheric pressure was introduced with an average speed of $0.146\,\mathrm{m/s}$. The microchannel was cut into a brass block and heated from three sides, while a plexiglass cover allowed visualisation from the fourth side. The temperature of the heated walls was reported as $102.1\,\mathrm{^\circ C}$. They observed bubble nucleation and growth along the microchannel and measured the bubble equivalent diameter versus time, see data in Fig.~\ref{mukherjee}(b). Additionally, they performed numerical simulations of the flow by using a level-set method, modelling a perfectly square microchannel cross-section, and initialising a steam bubble of diameter of $50\,\mathrm{\mu m}$ over a heated wall with a contact angle of $30^\circ$ (hydrophilic walls). Their numerical results are also included in Fig.~\ref{mukherjee}(b). 

The experimental setup is here emulated as a square microchannel ($D_h=229\,\mathrm{\mu m}$) of $5D_h$ length, with the solid region being disregarded. The channel is heated with a constant temperature of $102.1\,\mathrm{^\circ C}$ imposed on three sides while the fourth wall is set as adiabatic. Water at saturation temperature, $100\,\mathrm{^\circ C}$, enters the channel with a uniform velocity of $0.146\,\mathrm{m/s}$. No-slip and a static contact angle of $30^\circ$ are set at all walls. A vapour bubble of diameter of $50\,\mathrm{\mu m}$ is initialised at the heated wall, opposite the adiabatic one. Gravity is neglected owing to the small spatial scales. To obtain realistic velocity and temperature fields to be set as initial conditions at $t=0$, a preliminary liquid-only simulation is run till steady-state is achieved. The computational mesh is a structured orthogonal mesh made of uniform hexahedrons and grid spacing of $2.29\,\mathrm{\mu m}$. A similar mesh was used by \citet{mukherjee2011}; tests with a finer mesh did not show appreciable differences in the results. The simulation is evolved in time until the bubble nose approaches the outlet section of the microchannel.

The boiling solver is tested both in algebraic and geometric VOF modes. With both methods, tests are performed to assess the impact of the surface tension model with and without the density-correction term, and with different numbers of smoothing cycles for the interface curvature calculation (see Section~\ref{Sec:stModel}). Figure~\ref{mukherjee}(a) shows a snapshot of the flow dynamics little before the bubble reaches the outlet section. A thermal boundary layer develops over the heated walls as a result of the colder water coming through the inlet section. At the time instant displayed, the bubble has grown sufficiently to become elongated and dry patches form over the microchannel walls. The liquid evaporation rate is maximum at the solid-liquid-vapour contact line, where temperature is the highest. The bubble equivalent diameter versus time achieved with the different simulation setups are displayed in Fig.~\ref{mukherjee}(b). Without neither density-correction nor smoothing cycles, both isoAdvector and MULES (interFOAM) advection methods yield a significant overestimation of the bubble growth. This can be ascribed to the presence of parasitic currents related to errors in surface tension, which enhance convective flows near interfaces \cite{hardt}. isoAdvector seems to particularly suffer from parasitic currents, as it was previously observed in benchmark tests by \citet{gamet2020} and \citet{magnini2022a}, which is due to the sharper volume fraction changes at the interface resulting from the geometric VOF advection, which are detrimental to the interface curvature calculation when this is obtained via gradients of $\alpha$. The situation improves when the density-correction term in Eq.~\eqref{surfTens} is enabled, in particular from the instant when the bubble equivalent diameter grows above $D_h$ and the bubble becomes elongated. Spurious currents manifest in simulations of elongated bubbles in microchannels by generating artificial vortices in the liquid ahead the bubble nose \cite{abadie2015}, which explains the lower bubble growth rate trend observed in Fig.~\ref{mukherjee}(b) as density-correction is activated. The activation of smoothing cycles in the interface curvature calculation does not yield significant further changes in the bubble growth rate obtained with the MULES (though deviations appear at later stages), but it makes a significant difference when using isoAdvector. Since isoAdvector exhibits a sharper volume fraction field across the interface than MULES, a test with 4 smoothing cycles was also performed, leading to a further reduction of about $5\%$ of the bubble diameter achieved at the end of the simulation. 

Overall, the solver run with density-correction and isoAdvector mode with 4 smoothing cycles, or interFOAM mode with 2 smoothing cycles, yield results that agree well with the reference data of \citet{mukherjee2011}, with the geometric VOF exhibiting a bubble growth rate closer to \citet{mukherjee2011}'s simulation and the algebraic VOF following closely the experimental data points. It is worth noting that, though the present simulations were set similarly to the numerical setup of \citet{mukherjee2011}, a few impactful parameters from the experiment are unknown, e.g. the initial temperature field in the liquid at the instant of bubble nucleation, frequency of bubble nucleation and/or presence of other bubbles in the microchannel, actual temperature distribution over the boundary walls; these have a significant impact on the bubble growth rate, and thus deviations between experimental data and simulation results are expected.

\section{Results}
\label{Sec:results}

A systematic analysis of the effect of the channel aspect-ratio and thickness of the fins separating adjacent channels on the bubble dynamics and heat transfer performance for flow boiling in microchannels was conducted and the results are presented below, organised in subsections. First, the numerical setup is described in Sec.~\ref{Sec:setup}; this is followed by an analysis of the grid sensitivity of the numerical results  in Sec.~\ref{Sec:meshconv}. Then, the results obtained with liquid-only single-phase simulations are presented in Sec.~\ref{Sec:sp}, followed by the final Sec.~\ref{Sec:tp} where the flow boiling results are discussed.

\subsection{Simulation setup}
\label{Sec:setup}

\begin{figure}[t!]
\begin{center}
\includegraphics[width=100mm]{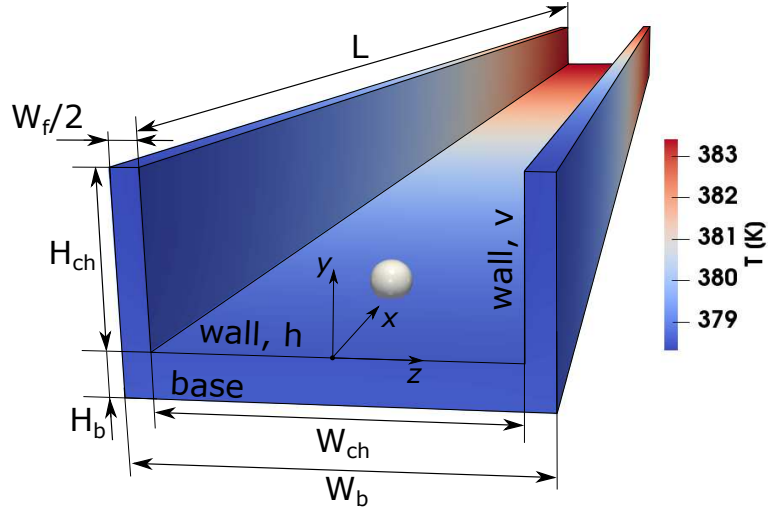}
\caption{Schematic of the simulation setup and notation used in this work. The microchannel has width $W_{ch}$ and height $H_{ch}$, with aspect-ratio $\epsilon=W_{ch}/H_{ch}$. The microchannel base has width and height $W_b$ and $H_b$, respectively. The vertical walls (fins) have thickness $W_f$ but, as symmetry boundary conditions are used at the sides of the domain, only half of the wall is modelled. At $t=0$, a vapour bubble is initialised at the centre of the bottom wall, at a distance of $1D_h$ from the channel inlet; the bubble has initial diameter of $0.2D_h$ and forms a contact angle of $30^\circ$ with the wall. Heat is applied to the bottom boundary of the base. The temperature profiles in both fluid and solid regions at $t=0$ are obtained with a preliminary liquid-only simulation run till steady-state is achieved. The image above refers to $\epsilon=2$ and $W_f=D_h/4$. }
\label{schematic}
\end{center}
\end{figure}

The geometrical configuration of multi-microchannel evaporators features several parallel microchannels etched in a block of conductive material, which are thus separated by solid walls (also called fins). The evaporator is typically heated from below, by placing its base surface in contact with a heater which applies a specified heat load. In experimental tests aimed at characterising two-phase heat transfer \cite{falsetti2018}, the top of the evaporator is often covered by a poorly conductive transparent material (glass, pyrex) to enable flow visualisation while limiting heat losses to the ambient.

In this numerical work, the geometrical configuration of a multi-microchannel evaporator is emulated by considering a single channel unit composed of one fluidic channel and the three connected solid wall regions at the two sides of the channel and below it; a schematic of the flow configuration and notation used in this work is provided in Fig.~\ref{schematic}. Symmetry boundary conditions are applied to the outer surface of the lateral walls to model the presence of adjacent channels. A constant and uniform heat flux is applied at the evaporator base. 

We consider microchannels of square and rectangular cross-sections, of constant hydraulic diameter $D_h=229\,\mathrm{\mu m}$ and length $L=20D_h$. The microchannel has width denoted as $W_{ch}$ and height $H_{ch}$, with the aspect-ratio defined as $\epsilon=W_{ch}/H_{ch}$; since the hydraulic diameter is maintained constant, the channel width and height are fully identified by $D_h$ and $\epsilon$: $W_{ch}=D_h(1+\epsilon)/2$,  $H_{ch}=D_h(1+\epsilon)/(2\epsilon)$. The microchannel base has width $W_b$ and thickness $H_b$, so that $W_b-W_{ch}=W_f$ identifies the thickness of the fins separating the channels. As boundary conditions, at the fluid inlet $(x=0,y,z)$ a liquid-only fully-developed laminar velocity profile of average speed $U_l$ is imposed, together with $T=T_{sat}$ and a zero-gradient condition for the pressure; the solid boundary at $(x=0,y,z)$ is set to adiabatic. At the fluid outlet $(x=L,y,z)$, zero-gradient conditions are set for both velocity and temperature with a uniform pressure value, and the solid boundary is adiabatic. On the bottom surface of the evaporator base, $(x,y=-H_b,z)$, a uniform heat flux is applied. The top boundary of the domain $(x,y=H_{ch},z)$ is adiabatic for both fluid $(|z| \leq  W_{ch}/2)$ and solid regions $(|z| \geq  W_{ch}/2)$. The three fluid-solid coupled boundaries are identified as vertical walls $(x,y,|z|=W_{ch}/2)$ and horizontal walls $(x,y=0,z)$; here, no-slip condition for the velocity, zero-gradient for the pressure and a static contact angle of $30^\circ$ for the volume fraction (hydrophilic walls) are imposed, as well as continuity of temperatures and heat fluxes (see Sec.~\ref{Sec:eqs}). Note that, since all the outer solid boundaries except the base are adiabatic, all the heat load applied through the base surface is dissipated via heat transfer to the fluid. The Bond number of the flow is $\mathrm{Bo}=\rho_l g D_h^2/\sigma < 0.01$, and thus gravitational forces are neglected. To achieve fully-developed velocity and temperature profiles as initial conditions for the two-phase simulation, a preliminary single-phase case with only liquid is run till steady-state. Figure~\ref{schematic} displays the solid walls of the channel, coloured with the steady temperature field, for a representative case. The results of the single-phase runs will be discussed in detail in Sec.~\ref{Sec:sp}, as they will be useful to rationalise the two-phase results. At $t=0$, the two-phase simulation starts with steady-state velocity and temperature fields, and a small spherical bubble of diameter $D_{eq}=0.2D_h$ is initialised near the inlet, sitting along the centreline over the bottom wall, with centroid of coodinates $(x=D_h,D_{eq}\cos(\pi/6)/2,0)$. Since the flow is expected to be symmetric to the $z=0$ plane, only half of the domain is simulated and symmetry boundary conditions are set at $(x,y,z=0)$. The two-phase simulation is run in time till the bubble reaches the outlet section of the channel, which is on the order of milliseconds. 

To perform this study, the channel hydraulic diameter is maintained constant and five different channel aspect-ratios are tested: $\epsilon=0.25,0.5,1,2,4$. For each channel aspect-ratio, four different values of the fin thickness are considered: $W_f=D_h, D_h/2, D_h/4, D_h/8$; these correspond to absolute fin thicknesses ranging from $28.6\,\mathrm{\mu m}$ to $229\,\mathrm{\mu m}$. The thickness of the base wall is maintained constant, $H_b=45.8\,\mathrm{\mu m}$. The working fluid is water at $p=1\,\mathrm{atm}$ and $T_{sat}=373.15\,\mathrm{K}$, with an inlet velocity of $U_l=0.146\,\mathrm{m/s}$ (mass flux $G=140\,\mathrm{kg/(m^2 s)}$); this corresponds to a value of the capillary number of $\mathrm{Ca}=\mu_l U_l/\sigma=0.0007$, as such very thin films and extended dry vapour patches are expected to form between the elongated bubble and the channel walls \cite{magnini2022a}. The Reynolds number of the flow is $\mathrm{Re}=\rho_l U_l D_h/\mu_l=100$, therefore the flow regime is laminar. A uniform heat flux of $q_b=100\,\mathrm{kW/m^2}$ is applied to the outer wall of the microchannel base, corresponding to a base power (for one single channel) ranging from $Q_b=0.08\,\mathrm{W}$ ($\epsilon=0.25$, $W_f=D_h/8$) to $Q_b=0.37\,\mathrm{W}$ ($\epsilon=4$, $W_f=D_h$) and a constant boiling number of $\mathrm{Bl}=q_b/(\rho_l U_l h_{lv})=0.0003$. The material composing the solid regions of the evaporator is copper, taken with constant properties $\rho_s=8940\,\mathrm{kg/m^3}$, $c_{p,s}=385\,\mathrm{J/(kg\,K)}$ and $\lambda_s=380\,\mathrm{W/(m\, K)}$.

Following on from the results presented in Sec.~\ref{Sec:validation}, for this study the boiling solver is set to run in isoAdvector mode (geometric VOF), with density-correction enabled and 4 smoothing cycles for the interface curvature calculation.

\subsection{Mesh convergence analysis}
\label{Sec:meshconv}

The computational domain is meshed with structured orthogonal meshes made of cubic hexahdrons. To identify the optimal mesh arrangement, a grid independence analysis was performed for the case with $\epsilon=1$ and $W_f=D_h/8$, with the domain length set to $L=10D_h$ to decrease the computational cost. Three different meshes are tested, with grid spacing of $\Delta=4.58\,\mathrm{\mu m}, 2.29\,\mathrm{\mu m}, 1.14\,\mathrm{\mu m}$, corresponding to $50, 100, 200$ mesh elements per hydraulic diameter. Overall, the three grids (with $L=10D_h$) have about 0.4, 3, and 24 million cells.

\begin{figure}[t!]
\begin{center}
\includegraphics[width=160mm]{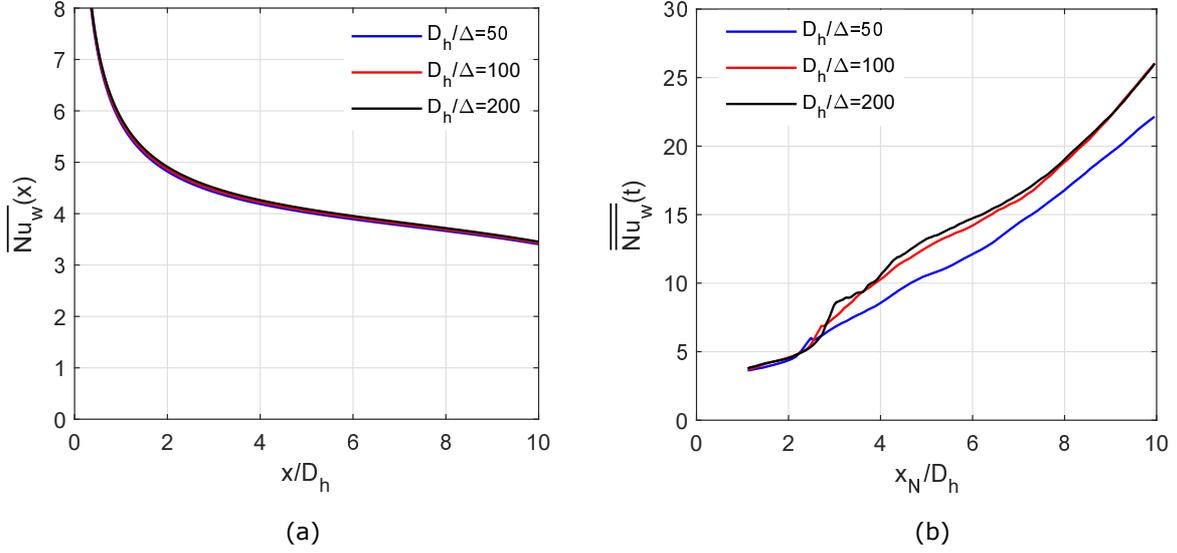}
\caption{Results of mesh convergence tests for (a) single-phase and (b) two-phase simulations. (a) Single-phase Nusselt number at steady-state, averaged around the cross-sectional perimeter, as a function of the streamwise coordinate; see definition in Eq.~\eqref{Eq:Nusp}. (b) Two-phase Nusselt number averaged over all three microchannel walls, see definition in Eq.~\eqref{Eq:Nutp}, plotted as a function of the location of the bubble nose.}
\label{meshconv}
\end{center}
\end{figure}

The grid independence analysis for the single-phase flow is first described. The Nusselt number calculated over the three fluid-solid wall boundaries, at steady-state, is considered. Local values of the Nusselt number are computed from the simulation data as:
\begin{equation}
\mathrm{Nu_w}(x,y,z)=\frac{q_w (x,y,z)}{[T_w (x,y,z)-T_{f}(x)]} \frac{D_h}{\lambda_l}
\label{Eq:Nu}
\end{equation}
where $q_w$ and $T_w$ are the heat flux and temperature on the horizontal and vertical channel walls in contact with the solid region; see schematic in Fig.~\ref{schematic}. The fluid temperature along the channel is evaluated by integrating an energy balance from $T=T_{sat}$ at the inlet:
\begin{equation}
T_f(x)=T_{sat}+\frac{2}{\rho_l c_{p,l} U_l W_{ch} H_{ch}} \int\limits_0^x \left[ \int\limits_{0}^{W_{ch}/2} q_w (x,y=0,z) \, dz + \int\limits_{0}^{H_{ch}} q_w (x,y,z=W_{ch}/2) \, dy\right] dx
\end{equation}
where the symmetry condition at $z=0$ has been used to simplify the integral. From Eq.~\eqref{Eq:Nu}, a cross-sectional average Nusselt number, $\overline{\mathrm{Nu_w}}(x)$, is obtained by averaging $\mathrm{Nu_w}(x,y,z)$ around the cross-sectional perimeter at a specific streamwise coordinate $x$:
\begin{equation}
\overline{\mathrm{Nu_w}}(x)=\frac{1}{W_{ch}/2+H_{ch}} \left[ \int\limits_{0}^{W_{ch}/2} \mathrm{Nu_w}(x,y=0,z) \, dz + \int\limits_{0}^{H_{ch}} \mathrm{Nu_w}(x,y,z=W_{ch}/2) \, dy\right]
\label{Eq:Nusp}
\end{equation}
The profiles of the average Nusselt number along the channel for the three meshes are plotted in Fig.~\ref{meshconv}(a). The three curves overlap and differences are below $1\%$, thus suggesting that all meshes are sufficient to properly resolve the single-phase flow. 

Next, the performances of the grids for the two-phase flow are presented. Owing to the bubble growth and elongation along the microchannel as time elapses, wall temperatures and heat fluxes vary significantly during the simulation and the Nusselt number is also a function of time, $\mathrm{Nu_w}(x,y,z,t)$; note that the Nusselt number for the two-phase flow is still calculated via Eq.~\eqref{Eq:Nu}, but replacing $T_f(x)$ with the constant $T_{sat}$. To compare the grids, a spatially-averaged Nusselt number is calculated during runtime, by averaging the Nusselt number over the three fluid-solid walls at each time instant:
\begin{equation}
\overline{\overline{\mathrm{Nu_w}}}(t)=\frac{1}{L(W_{ch}/2+H_{ch})} \int\limits_0^L \left[ \int\limits_{0}^{W_{ch}/2} \mathrm{Nu_w}(x,y=0,z,t) \, dz  + \int\limits_{0}^{H_{ch}} \mathrm{Nu_w}(x,y,z=W_{ch}/2,t) \, dy \right] dx
\label{Eq:Nutp}
\end{equation}
where $L$ is the channel length. The average Nusselt number versus time for the two-phase simulations and different meshes is plotted in Fig.~\ref{meshconv}(b). The time in the abscissa is expressed in terms of the location of the most downstream end of the bubble, or bubble nose, which is extracted at each time instant and is denoted as $x_N$. The medium and most refined meshes yield very similar bubble dynamics and the two curves of the Nusselt number versus bubble nose location overlap throughout the bubble growth and elongation process, whereas the coarsest mesh yields a $\sim 15\%$ smaller Nusselt number, due to a larger extension of dry vapour patches which can be ascribed to an insufficient resolution of the liquid film trapped between bubble and walls. Therefore, the mesh with 100 cells per hydraulic diameter was selected to run the analysis presented in the following sections.

Simulations were run on UK's Tier-1 supercomputer ARCHER2, which features computing nodes with two AMD EPYC 7742 2.25 GHz 64-core processors, thus making 128 cores and 256 GB of RAM per node. Simulations were run using two computing nodes and each two-phase run (with $L=20D_h$) required between 6,000 and 12,000 core-hours to complete.

\subsection{Single-phase results}
\label{Sec:sp}

In this section, the heat transfer results at steady-state obtained for single-phase simulations run with only liquid, in the range of aspect-ratios $\epsilon=0.25-4$ and fin thicknesses $W_f=D_h/8- D_h$, are presented. 

\subsubsection{Square channel and effect of fin width}

\begin{figure}[t!]
\begin{center}
\includegraphics[width=160mm]{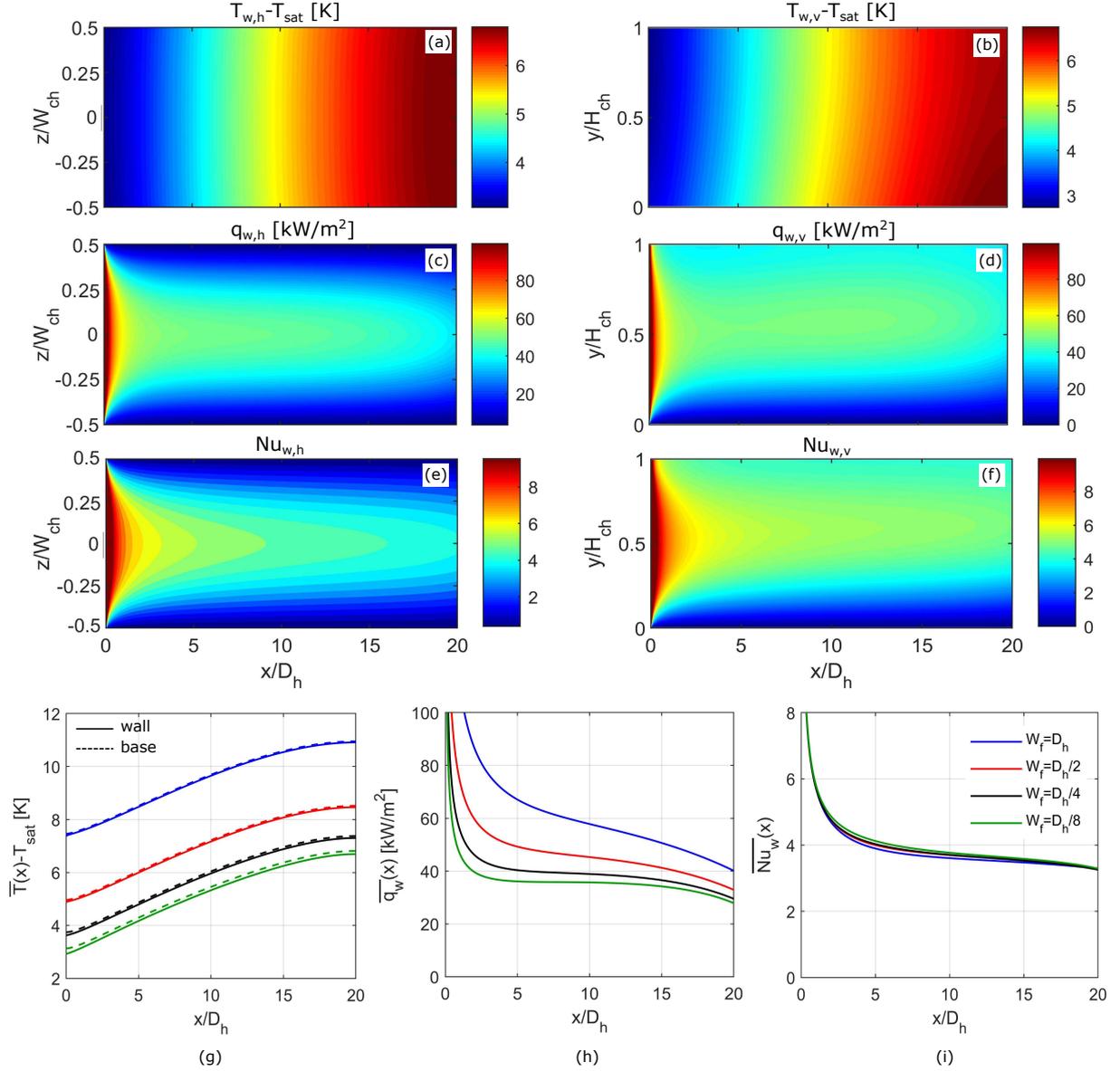}
\caption{Single-phase results for $\epsilon=1$ at steady-state. (a-f) Contours of (a,b) temperature, (c,d) wall heat flux, and (e,f) Nusselt number, over the (a,c,e) horizontal and (b,d,f) vertical walls of the channel, for $W_f=D_h/8$; figures not to scale. Plots of (g) wall and base temperatures, (h) wall heat flux, and (i) Nusselt number, averaged over the cross-stream direction, as a function of the streamwise coordinate, for different fin thicknesses. The legend in (i) applies also to (g) and (h).}
\label{sp_AR1_Wf16}
\end{center}
\end{figure}

Figure~\ref{sp_AR1_Wf16} illustrates contours of temperature, heat flux, and Nusselt number over the horizontal and vertical channel walls, and averaged over the cross-stream direction, for a representative case executed with $\epsilon=1$ and $W_f=D_h/8$. The heat flux and Nusselt number contours emphasise the development of thermal boundary layers over the coupled fluid-solid walls, $|z|/W_{ch}=0.5$ in Fig.~\ref{sp_AR1_Wf16}(c) and (e) and $y/H_{ch}=0$ in Fig.~\ref{sp_AR1_Wf16}(d) and (f), whereas smaller gradients appear nearby the channel top wall, $y/H_{ch}=1$ in Fig.~\ref{sp_AR1_Wf16}(d) and (f), owing to the adiabatic boundary condition. As a result of the different boundary conditions, the temperature distribution over the horizontal wall (Fig.~\ref{sp_AR1_Wf16}(a)) is uniform in the cross-stream direction whereas temperature decreases towards the top of the channel along the vertical wall (Fig.~\ref{sp_AR1_Wf16}(b)). Temperature, heat flux and Nusselt number are averaged around the cross-sectional perimeter as indicated in Eq.~\eqref{Eq:Nusp} and the resulting streamwise profiles are shown in Fig.~\ref{sp_AR1_Wf16}(g)-(i), for different fin widths. Figure~\ref{sp_AR1_Wf16}(g) includes also the cross-stream average of the base temperature, $T_b(x,z)=T(x,y=-H_b,z)$, calculated as:
\begin{equation}
\overline{\mathrm{T_b}}(x)=\frac{1}{W_b/2} \int\limits_{0}^{W_{b}/2} T_b(x,z) \, dz
\label{Eq:Tb}
\end{equation}
It can be seen that, though fluid enters the channel at $T_{sat}$, the average temperature of the microchannel wall at the inlet ($x/D_h=0$) is a few degrees higher than $T_{sat}$, due to the effect of axial heat conduction which redistributes heat towards the inlet region where the fluid is colder, and thus the temperature rise in the streamwise direction is milder than expected in the case of negligible conjugate heat transfer effects. Since the channel width $W_{ch}$ is maintained constant, thicker microchannel walls ($W_f$) yield larger wall and base temperatures as more heat is delivered to the fluid. This is expected because the base heat flux $q_b$ is maintained constant and a simple energy balance of the evaporator yields $q_b W_bL=\overline{\overline{q_w}} (2H_{ch}+W_{ch})L$, such that $\overline{\overline{q_w}}$ (average heat flux over all coupled fluid-solid walls) must increase as $W_b$ increases, in order to deliver the increased heat load to the fluid through the same exchange area $L(2H_{ch}+W_{ch})$. For example, the energy balance suggests $\overline{\overline{q_w}}=37.5\,\mathrm{kW/m^2}$ when $W_f=D_h/8$ and $\overline{\overline{q_w}}=66.7\,\mathrm{kW/m^2}$ when $W_f=D_h$, as confirmed by the wall heat flux magnitudes achieved in Fig.~\ref{sp_AR1_Wf16}(h). Owing to the high thermal conductivity of copper, wall and base temperatures are almost overlapping in Fig.~\ref{sp_AR1_Wf16}(g). A quick one-dimensional heat conduction calculation across the evaporator base suggests that the base-wall temperature difference should be $\approx q_b H_b/\lambda_s \approx 0.01\,\mathrm{K}$, thus validating this observation. Nonetheless, the base-wall temperature difference increases as the fin width is decreased, because the vertical wall temperature decreases owing to the lower thermal resistance of the fin. Despite the differences observed in boxes (g) and (h), the Nusselt number profiles in Fig.~\ref{sp_AR1_Wf16}(i) exhibit little dependence on the fin width, which is expected as the single-phase wall-fluid convective heat exchange is not directly impacted by the wall thickness.

\subsubsection{Effect of channel aspect-ratio}

\begin{figure}[t!]
\begin{center}
\includegraphics[width=155mm]{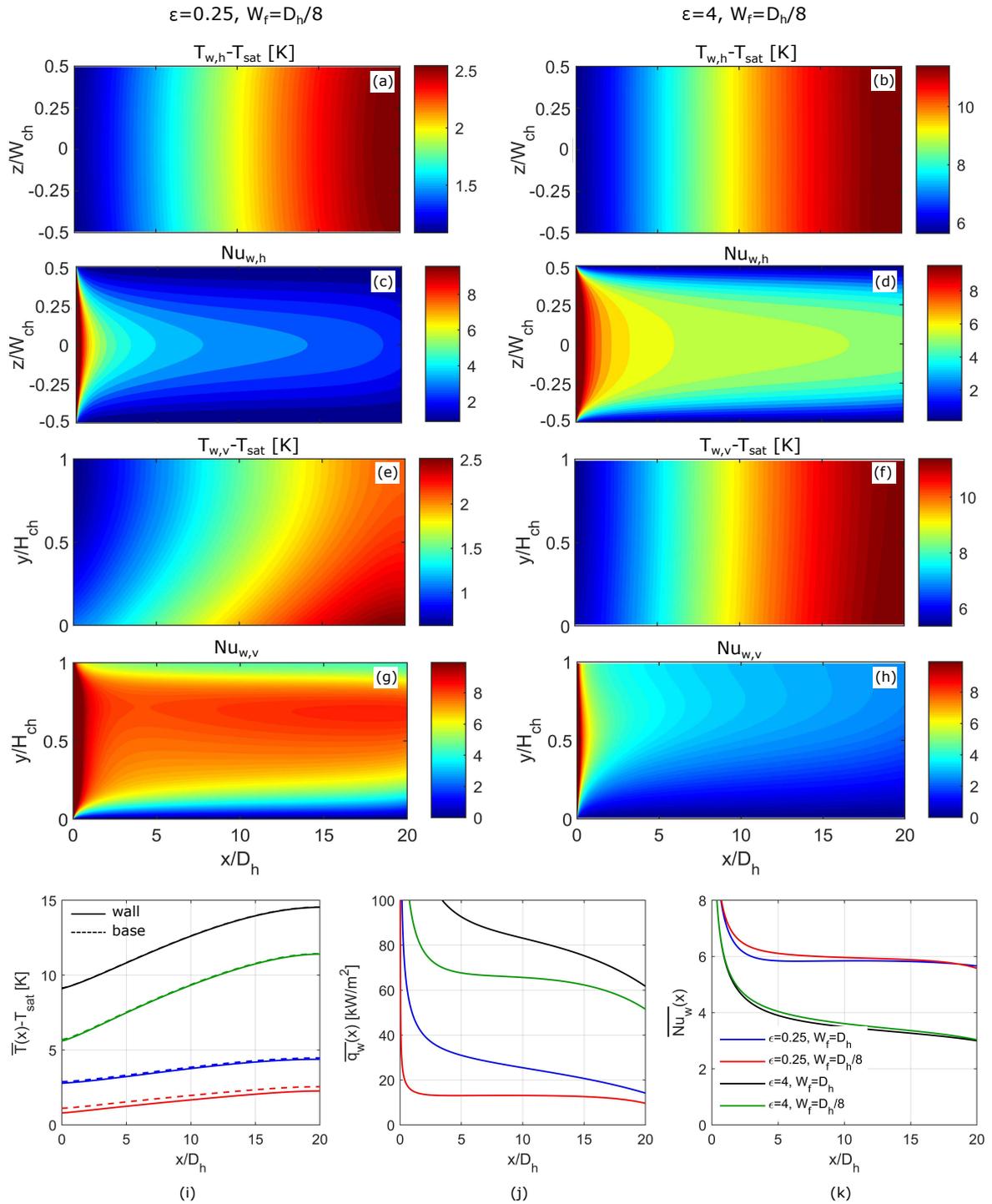}
\caption{Single-phase results for $\epsilon=0.25$ and $4$, at steady-state. The contours in (a,c,e,g) show (a,e) temperatures and (c,g) Nusselt numbers over (a,c) horizontal and (e,g) vertical walls for $\epsilon=0.25$ and $W_f=D_h/8$, while (b,d,f,h) show corresponding data for $\epsilon=4$ and $W_f=D_h/8$; figures are not to scale. The plots in (i,j,k) present wall and base temperatures, wall heat flux, and Nusselt number, averaged over the cross-stream direction, versus the streamwise coordinate, for the same aspect ratios and two selected values of the fin width. The legend in (k) applies also to (i) and (j).}
\label{sp_AR0p25-4}
\end{center}
\end{figure}

A comparison of the single-phase heat transfer performance for different channel aspect-ratios and same fin width is provided with Fig.~\ref{sp_AR0p25-4}. $H_{ch}>W_{ch}$ for $\epsilon=0.25$, and thus the shorter channel wall faces the base, whereas $W_{ch}>H_{ch}$ for $\epsilon=4$, where the longer wall is in contact with the evaporator base. The main differences between the two aspect-ratios are related to the different thicknesses of the thermal boundary layers developing over the channel walls. For $\epsilon=0.25$, the boundary layer develops quickly over the shorter horizontal wall and the local Nusselt number declines more rapidly, see Fig.~\ref{sp_AR0p25-4}(c), whereas heat convection over the horizontal wall remains effective along the microchannel for $\epsilon=4$, see Fig.~\ref{sp_AR0p25-4}(d), as the boundary layer over the wider wall developes more slowly. The situation is reversed over the vertical channel wall, where heat transfer is more effective for $\epsilon=0.25$, see Fig.~\ref{sp_AR0p25-4}(g) and (h), and the heightwise temperature distribution becomes less uniform (Fig.~\ref{sp_AR0p25-4}(e)). When comparing the contours of the Nusselt number over the wider wall, i.e. the vertical wall for $\epsilon=0.25$ (Fig.~\ref{sp_AR0p25-4}(g)) and the horizontal wall for $\epsilon=2$ (Fig.~\ref{sp_AR0p25-4}(d)), slightly better heat transfer is achieved for $\epsilon=0.25$, owing to the adiabatic top boundary of the channel (at $y/H_{ch}=1$) which allows for lower temperatures. 

The streamwise profiles of cross-stream averaged temperature and heat flux in Fig.~\ref{sp_AR0p25-4}(i) and (j) reveal that the wall and base temperatures are significantly lower for the smaller aspect-ratio channel, mainly because the heat applied through the evaporator base is smaller owing to the shorter channel width. For example, the same energy balance for the solid region appied above yields $\overline{\overline{q_w}}=13\,\mathrm{kW/m^2}$ when $\epsilon=0.25$ and $\overline{\overline{q_w}}=70\,\mathrm{kW/m^2}$ when $\epsilon=4$, both with  $W_f=D_h/8$. Since $\overline{\overline{q_w}}$ increases with increasing $\epsilon$, whereas the liquid mass flux is maintained constant, higher aspect-ratio channels lead to higher fluid temperatures, thereby resulting in larger evaporator temperatures.

The cross-stream average Nusselt number depicted in Fig.~\ref{sp_AR0p25-4}(k) confirms that the fluid-solid heat transfer is relatively insensitive to the fin width, but performances are substantially different for the two aspect-ratios despite the fact that the hydraulic diameter and the ratio between  the longer and shorter channel sides are the same (i.e. four) for both configurations. Counterintuitively, the configuration with $\epsilon=0.25$, where the shorter wall is in contact with the hot evaporator base, yields a significantly higher Nusselt number than that with the wider wall facing the heat source ($\epsilon=4$). This can be easily explained by inspection of the contours of the Nusselt number in Fig.~\ref{sp_AR0p25-4}(d) and (g). The magnitudes of the Nusselt number over the vertical wall for $\epsilon=0.25$ are comparable (or slightly higher) to those detected over the horizontal wall for $\epsilon=4$. However, owing to the three-side heating configuration, the heat transfer over the vertical wall contributes twice to the cross-stream average of $\mathrm{Nu_w}$, $\overline{\mathrm{Nu_w}}(x)=[W_{ch} \overline{\mathrm{Nu_{w,h}}}(x)+ 2H_{ch} \overline{\mathrm{Nu_{w,v}}}(x)]/(W_{ch}+2H_{ch})$,  and therefore the heat transfer performance over the vertical wall is twice as important as that over a horizontal wall of same size.  

\begin{figure}[t!]
\begin{center}
\includegraphics[width=160mm]{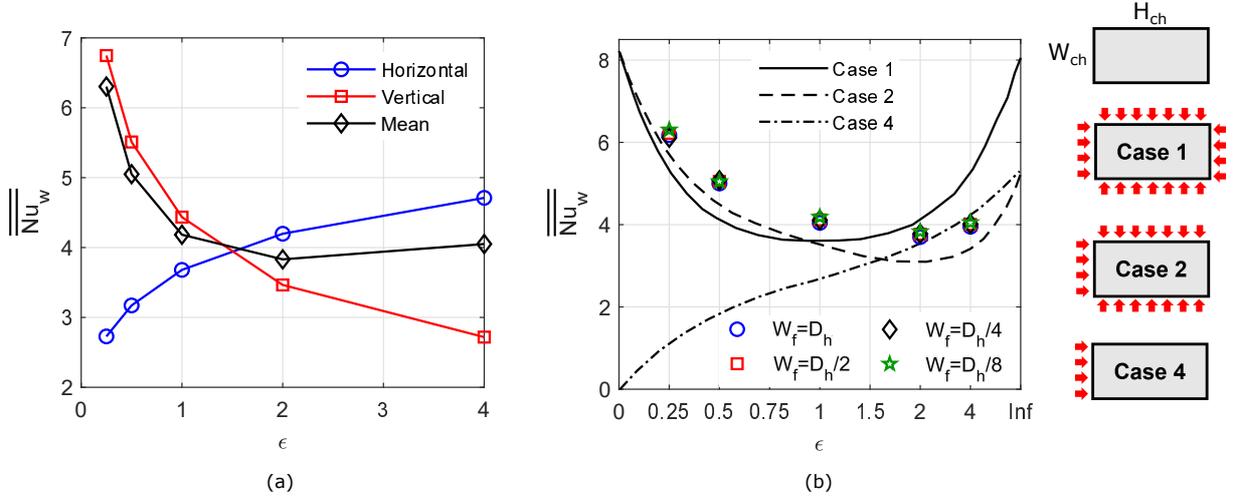}
\caption{Single-phase results for all aspect-ratios and fin widths, at steady-state. (a) Nusselt number, averaged over the vertical and horizontal walls, and averaged over all three microchannel walls (see Eq.~\eqref{Eq:Nutp}), for all aspect-ratios and $W_f=D_h/8$. (b) Nusselt number, averaged over all three microchannel walls, for different aspect-ratios and fin widths. Simulation data are compared with \citet{shahLondon} predictions for fully-developed laminar flow in a uniformly heated channel (Case 1), three-side heated channel (Case 2) and single-side heated channel (Case 4). }
\label{shahLondon}
\end{center}
\end{figure}

This aspect is investigated in detail in Fig.~\ref{shahLondon}, where the average Nusselt number over the entire microchannel heated walls (and separately for vertical and horizontal walls) is calculated using Eq.~\eqref{Eq:Nutp} and steady-state single-phase data. Though the Nusselt number over the horizontal wall increases as $\epsilon$ increases and that over the vertical wall decreases, and \textit{vice versa} as $\epsilon \rightarrow 0$, see Fig.~\ref{shahLondon}(a), the overall Nusselt number is larger at smaller aspect-ratios because, as mentioned above, the microchannel wall with better Nusselt number (vertical wall) contributes twice to the overall heat transfer. \citet{shahLondon} derived analytical solutions for hydrodynamically- and thermally-developed laminar flows in rectangular channels subject to nonuniform heating conditions and their results are displayed in Fig.~\ref{shahLondon}(b) together with the numerical data from the present work. While the \citet{shahLondon} results emphasize that there would be no difference in Nusselt numbers between $\epsilon=0.25$ and $4$ for uniform heating (Case 1), the heat transfer performance increases monotonically with $\epsilon$ for single-side heating (Case 4), whereas $\overline{\overline{\mathrm{Nu_w}}}$ exhibits a non-monotonic trend for the three-side heating identified as Case 2, with the Nusselt number being larger at smaller aspect-ratios in the range investigated in this study. Our numerical results follow very closely those for three-side heating, though the magnitudes obtained with the simulations are slightly larger as the flow is not thermally developed. The evident differences among the trends of $\overline{\overline{\mathrm{Nu_w}}}$ versus $\epsilon$ for different heating conditions observed in Fig.~\ref{shahLondon} emphasise the importance of including conjugate heat transfer in the microchannel model.

\subsubsection{Analysis of the entire single-phase database}
\label{Sec:model}

\begin{figure}[t!]
\begin{center}
\includegraphics[width=160mm]{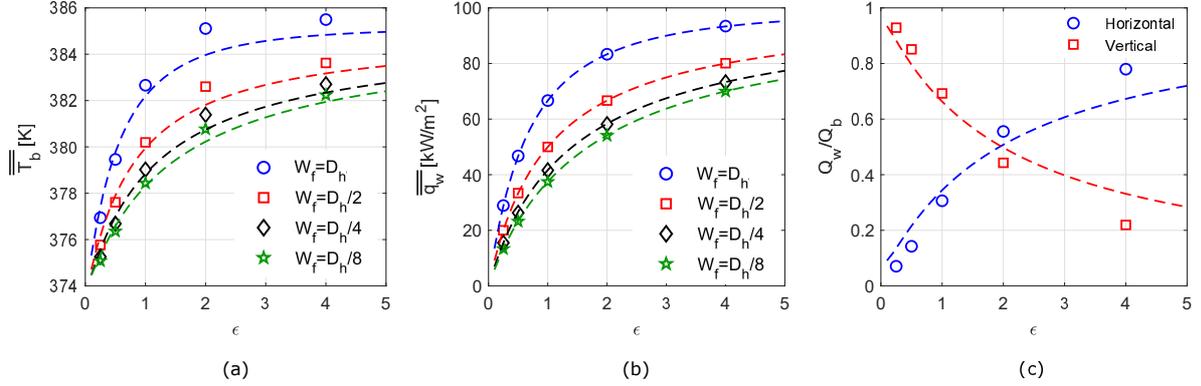}
\caption{Single-phase results for all aspect-ratios and fin widths, at steady-state. (a) Base temperature, averaged over the entire base surface, (b) wall heat flux averaged over all three microchannel walls, and (c) ratio of heat transferred through the three microchannel walls and heat applied at the base, for different aspect-ratio microchannels. The data in (c) refer to $W_f=D_h/8$. The dashed lines in all figures depict the predictions obtained with the heat transfer model; see Eqs.~\eqref{Eq:Tbmodel}, \eqref{Eq:qw} and \eqref{Eq:Qw}.  }
\label{singlePhase_TqNu}
\end{center}
\end{figure}

In order to compare all results for different aspect-ratios and fin widths, the whole numerical database of average base temperatures and wall heat fluxes is compiled in Fig.~\ref{singlePhase_TqNu}(a) and (b). Base temperatures and heat fluxes follow similar trends for the reasons explained above. Lower evaporator base temperatures are achieved with smaller channel aspect-ratios and thinner fin widths, because less heat per fluidic channel is applied to the evaporator. The differences induced by the width of the fins are less apparent at smaller aspect-ratios whereas they become more pronounced as $\epsilon$ increases. The plots in Fig.~\ref{singlePhase_TqNu} also include prediction curves obtained by means of a heat transfer model for the evaporator and microchannel. Using a calculation procedure usually applied to derive heat transfer coefficients from temperature measurements in microchannel flow boiling experiments \cite{alZaidi2021,szczukiewicz2014}, we consider a model of the evaporator composed of two resistences in series, the first related to heat conduction through the evaporator base, $R_b=H_b/(\lambda_s W_b L)$, and the second related to heat convection to the fluid. The latter is calculated as the result of two parallel thermal resistances, one over the horizontal and the other over the vertical channel walls. The resistance to convection at the horizontal wall is expressed as $R_{w,h}=(h_{w,h} W_{ch}L)^{-1}$, with $h_{w,h}$ being the heat transfer coefficient at the horizontal wall, that over the vertical wall is expressed using the fin efficiency, $R_{w,v}=(\eta h_{w,v} 2 H_{ch}L)^{-1}$, with $\eta=\text{tanh}(m H_{ch})/(mH_{ch})$ and $m=\sqrt{2h_{w,v}/(\lambda_s W_f)}$. The overall resistance to convection is then calculated as $R_w=R_{w,h} R_{w,v}/(R_{w,h}+R_{w,v})$ and the total resistance of the system is $R_{tot}=R_b+R_w$. Since the heat applied to the system via the evaporator base is known, $Q_b=q_b W_b L$, the average base temperature can be estimated as:
\begin{equation}
\overline{\overline{T_b}}=\overline{\overline{T_f}}+Q_b R_{tot}
\label{Eq:Tbmodel}
\end{equation}
where the average fluid temperature is calculated via an energy balance for the flow:
\begin{equation}
\overline{\overline{T_f}}=T_{sat}+\frac{1}{2}\frac{Q_b}{W_{ch} H_{ch} U_l \rho_l c_{p,l}}
\end{equation}
The predictions for the base temperature obtained with Eq.~\eqref{Eq:Tbmodel} are reported in Fig.~\ref{singlePhase_TqNu}(a) as dashed lines. The only unknowns in the model are the heat transfer coefficients $h_{w,h}$ and $h_{w,v}$, which are set by assuming a constant Nusselt number value of $4$. Estimations of the average wall heat flux can be obtained via the energy balance for the solid region used before:
\begin{equation}
\overline{\overline{q_w}}=\frac{Q_b}{(2H_{ch}+W_{ch})L}
\label{Eq:qw}
\end{equation}
which are plotted as dashed lines in Fig.~\ref{singlePhase_TqNu}(b). The heat transfer model captures very well the base temperature and wall heat flux trends versus aspect-ratio and fin width obtained with the numerical simulations, thus confirming that lower base temperatures can be achieved by decreasing the aspect-ratio of the microchannels. Figure~\ref{singlePhase_TqNu}(c) reports the ratio of heat transferred through the horizontal or vertical wall and the base heat load, when varying the aspect-ratio for $W_f=D_h/8$. The predictions displayed as dashed lines are obtained as:
\begin{equation}
Q_{w,h}=\frac{\overline{\overline{T_w}}-\overline{\overline{T_f}}}{R_{w,h}}, \quad Q_{w,v}=\frac{\overline{\overline{T_w}}-\overline{\overline{T_f}}}{R_{w,v}}
\label{Eq:Qw}
\end{equation}
where $\overline{\overline{T_w}}=\overline{\overline{T_b}}-q_bH_b/\lambda_s$; note that the model calculates an average wall temperature and does not discriminate between horizontal and vertical wall temperatures. At smaller aspect-ratios, most of the heat is transferred through the vertical wall (above 90\% for $\epsilon=0.25$). Even for the square channel ($\epsilon=1$), the vertical wall contributes to about 70\% of the heat dissipation to the fluid, as the vertical wall counts twice on the overall heat transfer balance owing to the three-side heating configuration. At larger aspect-ratios, heat transfer through the horizontal wall takes over, though the vertical wall still provides a considerable contribution, which explains the asymmetry on the profiles of the Nusselt number around $\epsilon=1$ in Fig.~\ref{shahLondon}.

\subsection{Two-phase results}
\label{Sec:tp}

\begin{figure}[t!]
\begin{center}
\includegraphics[width=165mm]{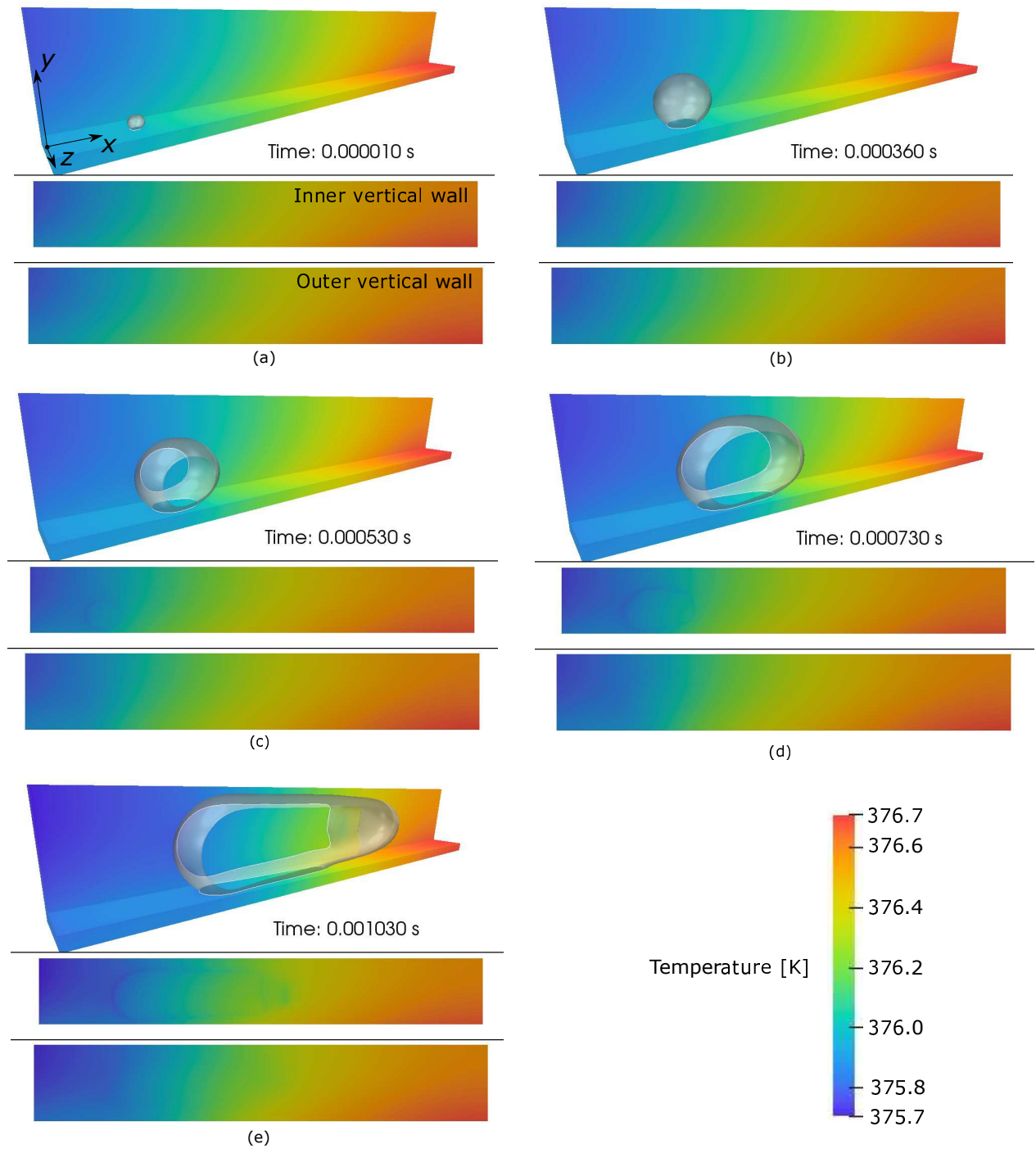}
\caption{Snapshots of bubble growth and temperature field evolution over time for a simulation run with $\epsilon=0.5$ and $W_f=D_h/2$. A shorter channel ($L=10D_h$) was used for the sake of visualisation. Within each box, the top image shows the bubble surface in grey and slight transparency, with the contact lines over the horizontal and left vertical walls highlighted in white. The right vertical wall of the channel is clipped to enable visualisation. The middle image shows the temperature field on the inner surface of the vertical wall, in contact with fluid. The bottom image shows the temperature field on the outer surface, where symmetry boundary conditions are applied. }
\label{snapshots}
\end{center}
\end{figure}

This section presents the results obtained with two-phase flow simulations, in the range of aspect-ratios $\epsilon=0.25-4$ and fin thicknesses $W_f=D_h/8- D_h$. To illustrate the bubble dynamics and heat transfer for a representative case, Fig.~\ref{snapshots} shows snapshots of the bubble growth and resulting microchannel temperatures as time elapses, for $\epsilon=0.5$ and $W_f=D_h/2$. At $t=0$, the temperature field corresponds to the single-phase steady-state solution. At the onset of the two-phase flow, the bubble grows expanding the dry vapour region over the bottom surface of the channel and, as its diameter reaches $W_{ch}$, a contact line is formed over the vertical wall. The inner vertical wall responds instantaneously to the presence of the contact line, see the central box in Fig.~\ref{snapshots}(c), as manifested by the colder nearly-circular spot revealed by the temperature contours, whereas this is not seen on the outer wall due to heat spreading. As the bubble grows further, it elongates along the channel and the dry vapour regions expand, leaving thick liquid lobes at the channel corners and thin liquid layers between the bubble nose and the downstream ends of the contact lines. As the bubble and contact lines progress along the channel, the wall temperature reduces, exhibiting larger gradients in the proximity of the contact lines. As the channel wall comes in contact with vapour in the dryout regions, the wall temperature tends to become more uniform and to increase due to the less efficient heat convection, though this happens with some delay due to the thermal inertia of the solid.

\subsubsection{Square channel and effect of fin width}

\begin{figure}[b!]
\begin{center}
\includegraphics[width=160mm]{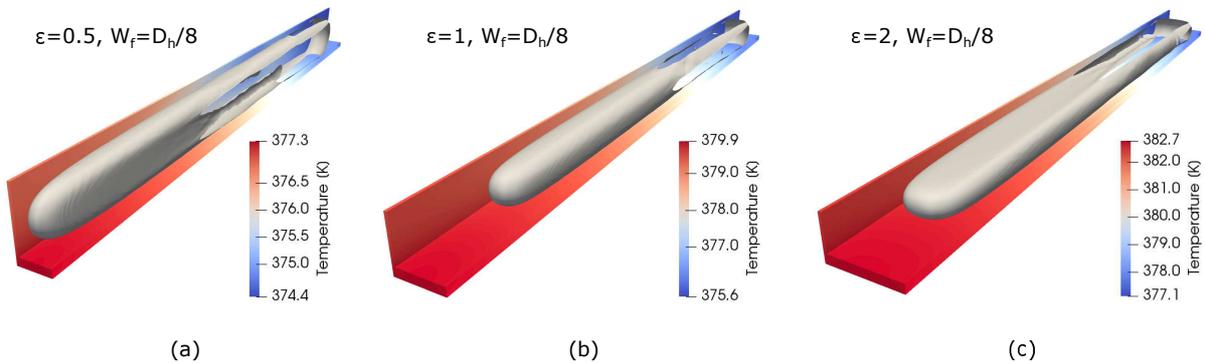}
\caption{Snapshots of bubble surface and temperature fields over the evaporators walls as the bubble approaches the outlet section, for selected channel aspect-ratios. One of the two vertical walls of the channel is clipped to enable visualisation.}
\label{shapes}
\end{center}
\end{figure}

\begin{figure}[t!]
\begin{center}
\includegraphics[width=160mm]{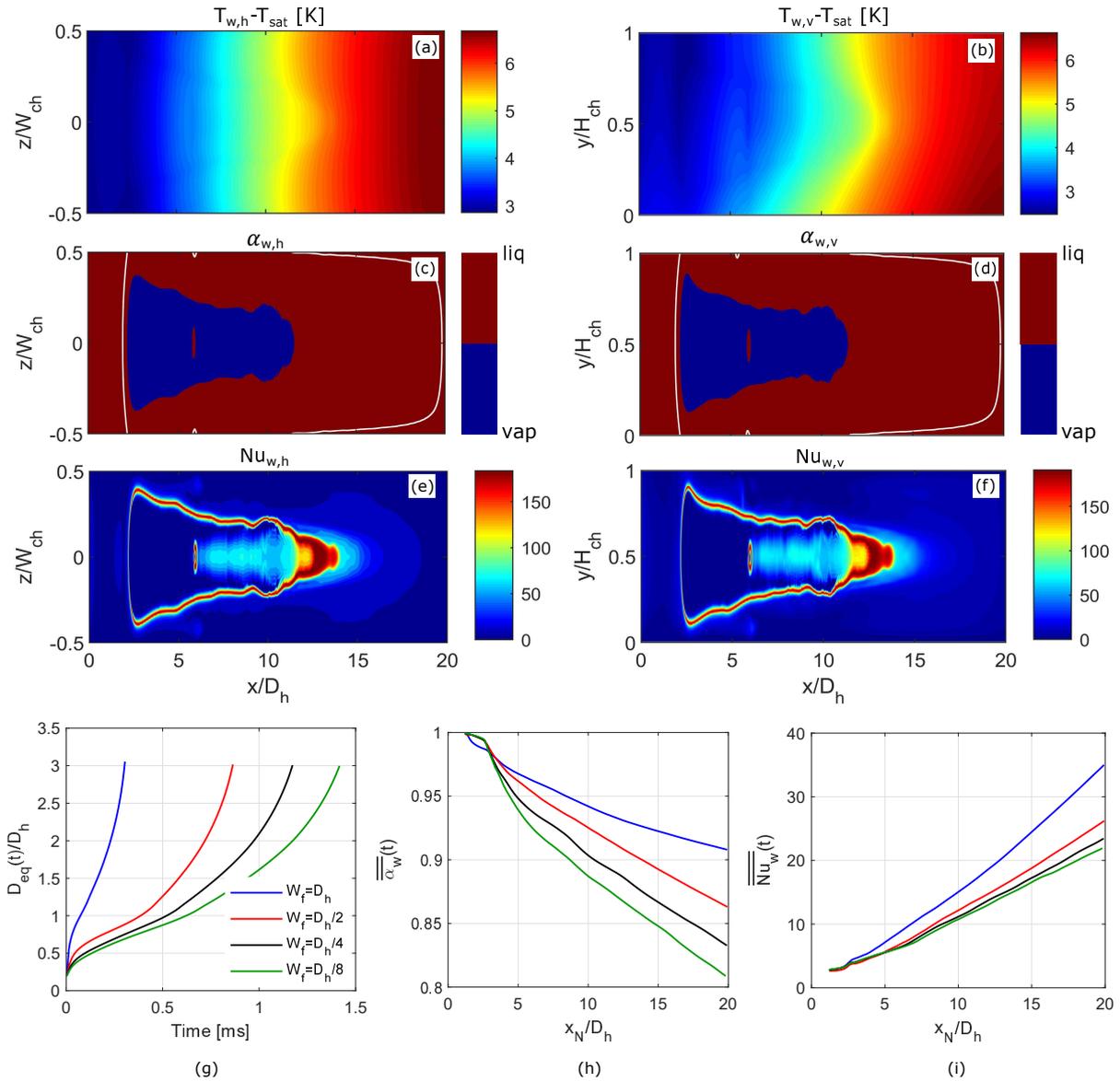}
\caption{Two-phase results for $\epsilon=1$. (a-f) Contours of (a,b) temperature, (c,d) liquid and vapour fraction, and (e,f) Nusselt number, over the (a,c,e) horizontal and (b,d,f) vertical walls of the channel, for $W_f=D_h/8$ at $t=t_{end}$; figures not to scale. The white lines in (c,d) identify the bubble profiles extracted on the (c) $y=H_{ch}/2$ and (d) $z=0$ planes. (g) Bubble diameter versus time. (h) Wet area fraction and (i) Nusselt number, averaged over all three microchannel walls, plotted as a function of the location of the bubble nose along the channel. The legend in (g) applies also to (h) and (i). }
\label{tp_AR1_Wf16}
\end{center}
\end{figure}

We begin the two-phase analysis with discussing the results obtained for a reference case run with a square channel ($\epsilon=1$). Figure~\ref{shapes} presents snapshots of the bubble surface and corresponding evaporator temperature fields for $\epsilon=0.5,1,2$. For the square channel, the bubble front has a cylindrical shape which flattens at the centre of the channel walls. Dry vapour patches are formed near the tail of the bubble, at the centre of the channel walls. The contours of temperature, wet fraction and Nusselt number over the microchannel walls detected at $t=t_{end}$, when the bubble nose reaches the outlet section, for $W_f=D_h/8$, are shown in Fig.~\ref{tp_AR1_Wf16}(a)-(f). At this stage, the bubble is elongated and dry vapour patches are clearly visible on the channel walls in Fig.~\ref{tp_AR1_Wf16}(c) and (d). Note that the wall dry regions are identified as boundary faces where the face-interpolated liquid volume fraction is $\alpha_w<0.5$, with $\alpha_w>0.5$ identifying wet regions. Therefore, the boundaries between the blue and red regions in Fig.~\ref{tp_AR1_Wf16}(c) and (d) do not have to be interpreted as sharp liquid-vapour boundaries. 
As expected, the largest values of the Nusselt number coincide with contact lines and very thin liquid films. 
The temperature field over the walls, Fig.~\ref{tp_AR1_Wf16}(a) and (b), is very sensitive to the two-phase flow and the walls progressively cool down as the contact line sweeps them. Figure~\ref{tp_AR1_Wf16}(g)-(i) reports the bubble equivalent diameter and the spatial average of wall wet fraction and Nusselt number, both calculated as indicated in Eq.~\eqref{Eq:Nutp}, as time elapses for all the fin widths tested. Time in the abscissa is expressed in terms of the location of the bubble nose $x_N$. The bubble growth rate increases with the fin width, as a result of the larger amount of heat stored by the solid regions during the previous single-phase stage, now dissipated in the form of latent heat. As the bubble grows, dry vapour patches are formed and expand over the walls, explaining the descending trends of $\overline{\overline{\alpha_w}}$ with increasing time. For larger $W_f$, the bubble nose propagates faster along the channel because the bubble grows more rapidly, and thus thicker liquid films are left at the channel walls, as expected from traditional lubrication theory \cite{bretherton1961}. Thicker films are less prone to dewetting and dryout, thus mitigating the expansion of dry regions and explaining the ascending trends of $\overline{\overline{\alpha_w}}$ with $W_f$ observed in Fig.~\ref{tp_AR1_Wf16}(h). The liquid film coverage of the wall explains also the trends of the Nusselt number reported in Fig.~\ref{tp_AR1_Wf16}(i). $\overline{\overline{\mathrm{Nu_w}}}$ increases over time as the channel wall becomes covered by a thin liquid film and both contact line and thin film evaporation contribute to cool down the wall. However, the Nusselt number increases at a larger rate for a thicker fin width, owing to the larger fraction of channel wall that remains covered by a liquid film. Note that the Nusselt number does not reach any steady-state or steady-periodic value, because multiple nucleation cycles  \cite{magnini2016b} and simulations over $100\times$ larger time-scales (order of $10^{-1}\,\mathrm{s}$) would be necessary for the temperature field within the solid region to achieve a statistically steady regime.

\subsubsection{Effect of channel aspect-ratio}

\begin{figure}[t!]
\begin{center}
\includegraphics[width=151mm]{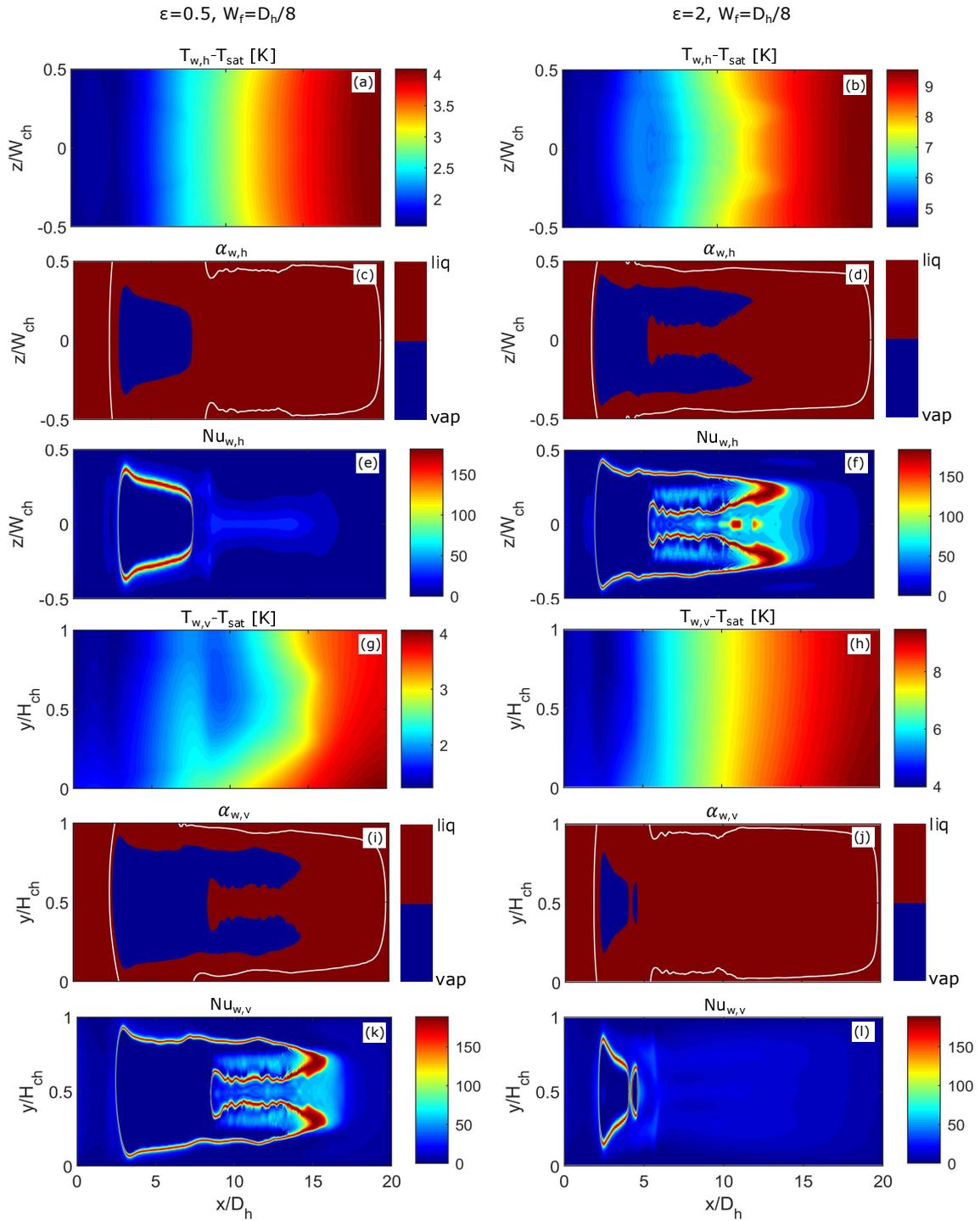}
\caption{Two-phase results for $\epsilon=0.5$ and $2$, both with $W_f=D_h/8$, at $t=t_{end}$. The contours in (a,c,e,g,i,k) show (a,g) temperatures, (c,i) liquid and vapour fractions, and (e,k) Nusselt numbers over (a,c,e) horizontal and (g,i,k) vertical walls for $\epsilon=0.5$, while (b,d,f,h,j,l) illustrate the corresponding contours for $\epsilon=2$; figures are not to scale. The white lines in (c,d,i,j) identify the bubble profiles extracted on the (c,d) $y=H_{ch}/2$ and (i,j) $z=0$ planes. }
\label{tp_AR0p5-2}
\end{center}
\end{figure}

Next, the results for two different aspect-ratios, $\epsilon=0.5$ and $2$, are illustrated. Figure~\ref{shapes} shows bubble shapes and corresponding evaporator temperature fields. Extended dry regions form over the wider channel walls, whereas much smaller dry regions appear over the shorter walls; these features are discussed below. Contours of wall temperatures, liquid and vapour fraction, and Nusselt numbers for $\epsilon=0.5,2$ at same $W_f=D_h/8$ are depicted in Fig.~\ref{tp_AR0p5-2}. The contours of the dry and wet area fraction, Fig.~\ref{tp_AR0p5-2}(c), (d), (i) and (j), confirm that larger dry regions develop over wider walls, as observed in Fig.~\ref{shapes}. This is a result of surface tension forces, that arrange the bubble cross-section into circular arcs facing the shorter walls, where a thicker liquid film develops, while the liquid-vapour interface is rather flat along the wider walls, where a thinner liquid film is left \cite{magnini2022a,magnini2020a}. Since the capillary number of the flow is small, $\mathrm{Ca}=0.0007$, the liquid film partially dries over both horizontal and vertical walls, but the dry area fraction is smaller over the shorter wall owing to the thicker film. It is interesting to note both in Fig.~\ref{tp_AR0p5-2}(d) and (i) that the dry region at the upstream end of the contact line develops into two axial dry streaks at its downstream end. This is due to the effect of capillary forces on the thin film covering the larger wall, that create a saddle-like film profile on the channel cross-section, which is thicker at the channel centre and exhibits a dimple at the matching point with the static meniscus at the side \cite{magnini2022a,hazel2002}. Dryout initiates in coincidence with this interfacial dimple where the film is the thinnest \cite{khodaparast2017,khodaparast2018}, as observed in Fig.~\ref{tp_AR0p5-2}(d) and (i) and, owing to the hydrophilic walls, a narrow liquid ligament still exists along the wall centreline between the two dry regions, although it eventually evaporates and dries out, thus leaving an extended dry region. Dry patches are more extended for $\epsilon=0.5$, because the bubble grows more slowly due to the lower wall temperature. The white lines included in Fig.~\ref{tp_AR0p5-2}(c), (d), (i), and (j), indicate the bubble profiles extracted along channel centreplanes, and reveal that the liquid ligament (e.g., Fig.~\ref{tp_AR0p5-2}(c), $x/D_h=9-14$) is thicker than the liquid film established in the downstream fully-wetted region ($x/D_h>14$). This is due to the fact that surface tension rearranges the cross-sectional profile of the liquid-vapour interface from the fully-wetted to the partially-dry region. In the downstream wet region, the film thickness is determined by the meniscus at the channel corners \cite{magnini2022a}, whereas the thickness of the liquid ligament is determined by the contact angle.

\begin{figure}[b!]
\begin{center}
\includegraphics[width=160mm]{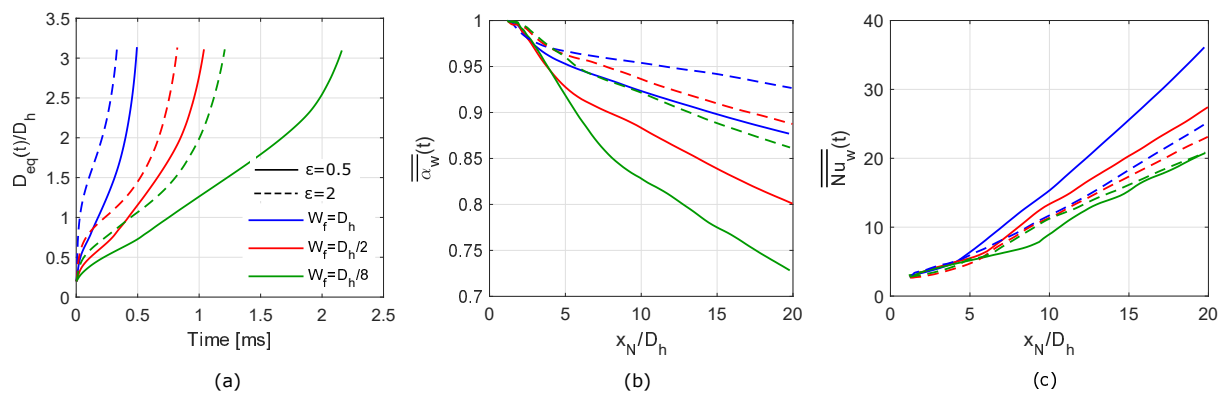}
\caption{Two-phase results for $\epsilon=0.5$ and $2$ and selected values of fin widths.  (a) Bubble diameter versus time, (b) wet area fraction and (c) Nusselt number, averaged over all three microchannel walls, plotted as a function of the location of the bubble nose along the channel.  The legend in (a) applies also to (b) and (c).}
\label{tp_AR0p5-2-mat}
\end{center}
\end{figure}

 The temperature and Nusselt number contours in Fig.~\ref{tp_AR0p5-2} are a result of the liquid film and contact line distribution along shorter and larger channel walls, as such the contours of $\overline{\overline{T_{w,h}}}$ (or $\overline{\overline{T_{w,v}}}$) for $\epsilon=0.5$ are qualitatively similar to those of $\overline{\overline{T_{w,v}}}$ (or $\overline{\overline{T_{w,h}}}$) for $\epsilon=2$, with the Nusselt number following analogous trends. The only apparent differences are on the contours of temperature over the larger wall, Fig.~\ref{tp_AR0p5-2}(b) and (g), where the vertical wall for $\epsilon=0.5$ restarts heating up in the dry vapour region. 
 
 A quantitative comparison of the results for $\epsilon=0.5$ and $2$ and different fin widths is presented in Fig.~\ref{tp_AR0p5-2-mat}. The vapour bubble grows more rapidly for $\epsilon=2$ because the microchannel walls are warmer, as a result of the previous single-phase steady-state temperature field, see Fig.~\ref{sp_AR0p25-4}. As a consequence, the larger aspect-ratio channel exhibits smaller dry vapour patches and larger wet area fraction (Fig.~\ref{tp_AR0p5-2-mat}(b)), although the curves of $\overline{\overline{\alpha_{w}}}$ for the two aspect-ratios converge as the fin width increases. Although the Nusselt numbers over the wider walls (vertical wall for $\epsilon=0.5$, horizontal for $\epsilon=2$) are of comparable magnitude, Fig.~\ref{tp_AR0p5-2-mat}(c) shows that higher average Nusselt numbers are achieved for a smaller aspect-ratio, since the vertical wall counts twice towards $\overline{\overline{\mathrm{Nu_{w}}}}$. The situation is reversed for $W_f=D_h/8$, see the green curves in Fig.~\ref{tp_AR0p5-2-mat}(c), which is the case illustrated in Fig.~\ref{tp_AR0p5-2}, because the bubble grows much faster for the larger aspect-ratio channel and therefore the wet area fraction for $\epsilon=2$ is considerably higher than that for $\epsilon=0.5$.

\subsubsection{Conjugate heat transfer analysis}

\begin{figure}[b!]
\begin{center}
\includegraphics[width=160mm]{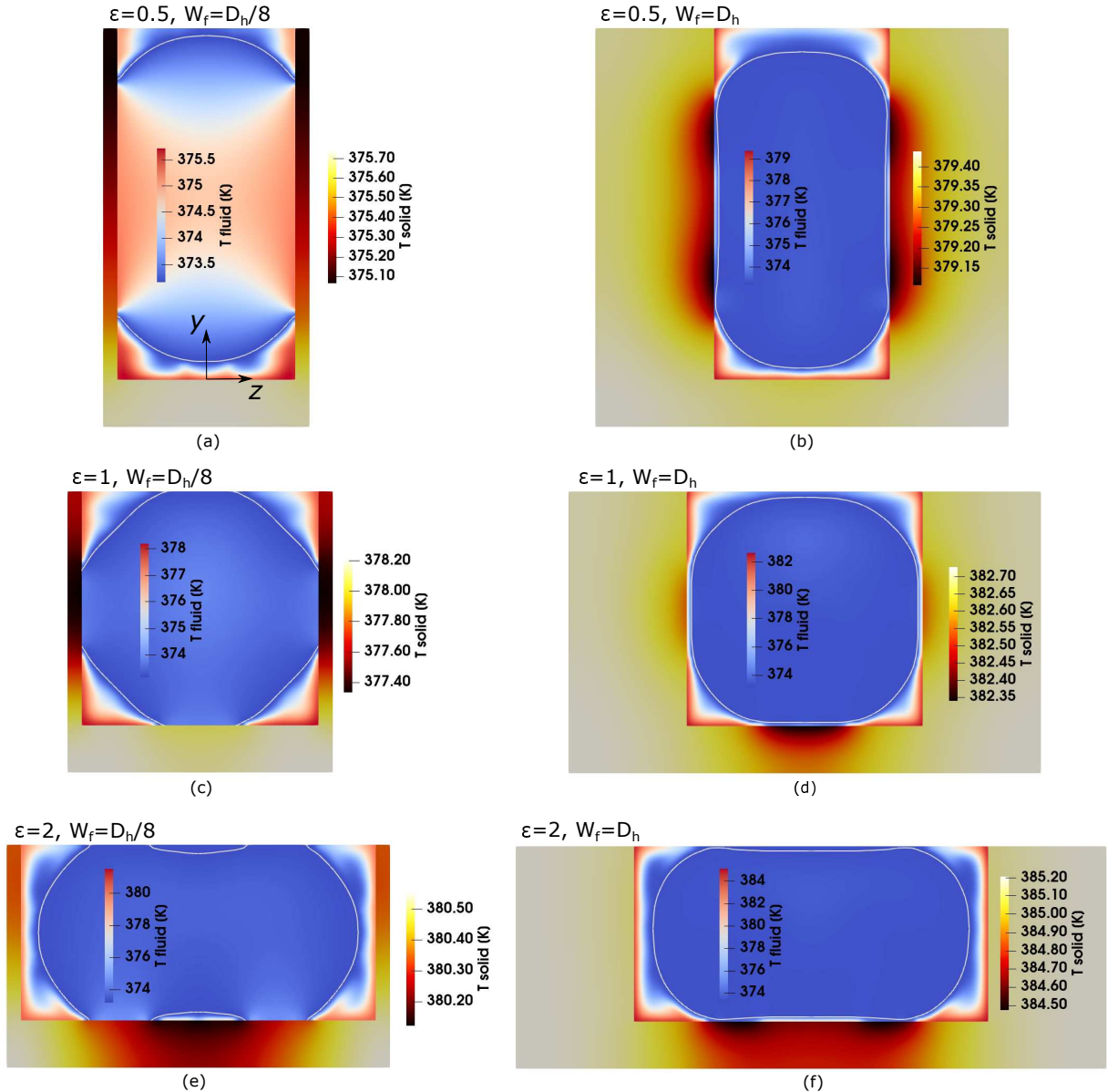}
\caption{Two-phase temperature fields on the channel cross-section ($x=10D_h$) for different aspect-ratios and fin widths. Two different colour scales are used for the temperature fields in the fluid and solid regions to better reveal local gradients. The white lines in the fluid regions identify the liquid-vapour interface on the channel cross-section. }
\label{crossSections}
\end{center}
\end{figure}

To better reveal the nonuniform temperature distribution within the solid regions, Fig.~\ref{crossSections} shows cross-sectional temperature fields within both solid and fluid, extracted half-way along the evaporator, for $\epsilon=0.5,1,2$, and two selected values of fin thicknesses. The images also include the profile of the liquid-vapour interface on the cross-section. When $W_f=D_h/8$, Fig.~\ref{crossSections}(a), (c) and (e), the bubble grows at a slower rate than $W_f=D_h$ and the liquid film over the wider walls is dry at the displayed cross-section, except for $\epsilon=2$ where the bubble grows faster owing to the warmer wall; thus, a narrow liquid ligament is left between the dry regions at the sides, as observed in Fig.~\ref{tp_AR0p5-2}(d). Thicker liquid films always cover the shorter walls at the selected cross-sections. When $W_f=D_h$, Fig.~\ref{crossSections}(b), (d) and (f), the bubble grows more rapidly as the channel walls are warmer and thin liquid films cover the channel walls for all aspect-ratios; the saddle-like shape of the liquid-vapour interface over the wider walls is clearly evident in Fig.~\ref{crossSections}(b) and (f), with minimum film thickness regions appearing at the matching point between the static meniscus at the corner and the thin film at the wall centre. By inspection of the temperature field in the solid and fluid regions, it is evident that the temperature is more uniform in the solid due to its much larger thermal conductivity. Larger temperatures are measured in both the fluid and solid as the aspect-ratio increases due to the increased heat load through the evaporator base. The impact of contact lines and thin films on the solid wall temperature are apparent. Temperature is the lowest near contact lines and very thin films, e.g. in the correspondence of the interface dimples for $\epsilon=0.5,2$, while temperatures are higher where thicker films cover the walls. Wall temperatures remain relatively low in dry vapour regions, due to the thermal inertia of the solid. Negligible temperature variation is observed in the horizontal direction across the evaporator fins for thin walls ($W_f=D_h/8$), whereas  horizontally-oriented gradients become evident for thicker walls ($W_f=D_h$), depending on the local liquid film morphology. 

\subsubsection{Base temperatures and analysis of the entire two-phase database}

\begin{figure}[t!]
\begin{center}
\includegraphics[width=160mm]{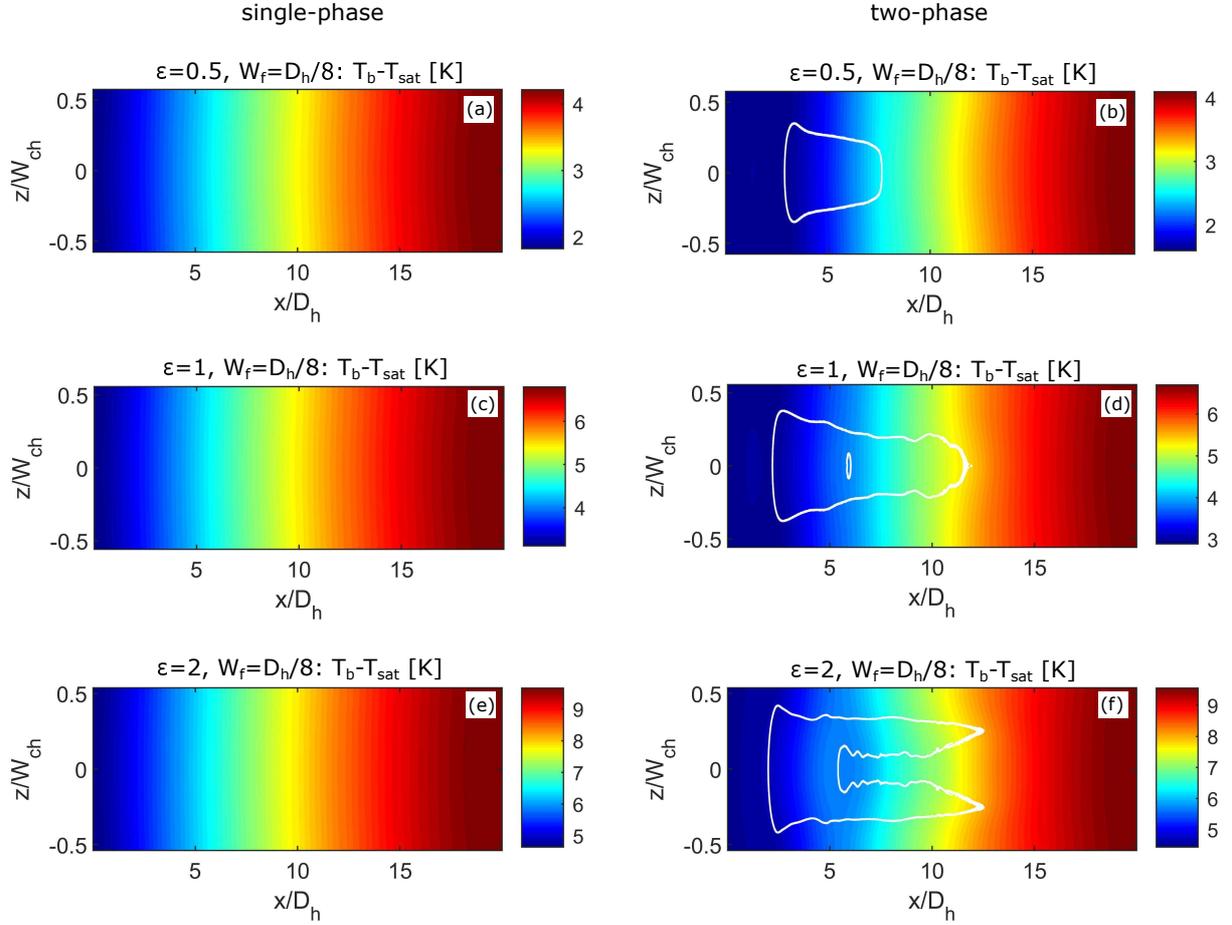}
\caption{Comparison of (a,c,e) single-phase and (b,d,f) two-phase base temperatures, for $\epsilon=0.5,\,1,\,2$, and $W_f=D_h/8$; figures are not to scale. The single-phase contours are exctracted at steady-state, whereas the two-phase results are taken at $t=t_{end}$. The white profiles in the two-phase temperature contours show the contact line formed by the bubble over the horizontal wall at $t=t_{end}$. }
\label{Tbase}
\end{center}
\end{figure}

The contours of temperature of the evaporator base for both single-phase (steady-state) and two-phase ($t=t_{end}$) simulations, $\epsilon=0.5,1,2$ and $W_f=D_h/8$ are presented in Fig.~\ref{Tbase}. As discussed previously, temperatures increase when increasing $\epsilon$ and more heat must be dissipated by the fluid. For single-phase flow, the spanwise distribution of temperature is uniform. The formation of a contact line and an evaporating film which are both advancing along the channel cool down the walls and this effect is propagated to the evaporator base via heat conduction through the solid. The cooling of the evaporator base is manifested by the isolines of temperature shifting downstream in Fig.~\ref{Tbase}(b), (d) and (f). The spanwise temperature profile for the two-phase flow stays rather uniform for $\epsilon=0.5$ and 1 owing to the limited extension of the evaporating film, whereas for $\epsilon=2$ the base centreline appears colder than the corners due to the wide liquid film established over the horizontal wall; however, part of this film eventually dries out and as time elapses the less effective solid-vapour heat convection may lead to a local increase of temperatures.

\begin{figure}[t!]
\begin{center}
\includegraphics[width=160mm]{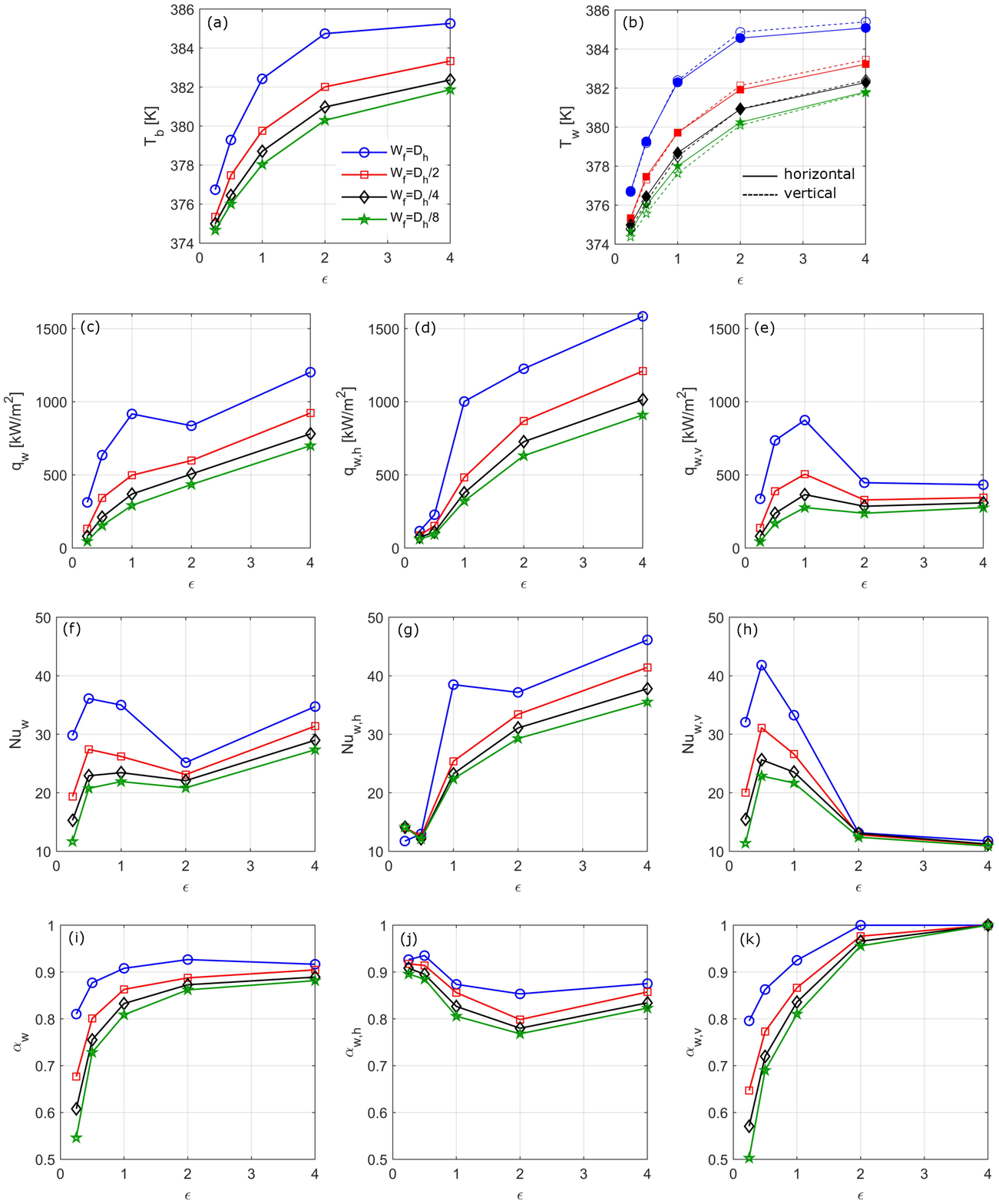}
\caption{Two-phase results for all aspect-ratios and fin widths, extracted at $t=t_{end}$. Average (a) base and (b) wall temperatures. (c,d,e) Heat fluxes, averaged over (c) the three microchannel walls, (d) horizontal-only and (e) vertical-only walls. (f,g,h) Nusselt numbers, averaged over (f) the three microchannel walls, (g) horizontal-only and (h) vertical-only walls. (i,j,k) Wet area fractions, averaged over (i) the three microchannel walls, (j) horizontal-only and (k) vertical-only walls. The double-overline notation is here dropped for convenience.}
\label{twoPhase_TqNu}
\end{center}
\end{figure}

To compare all the two-phase results at varying aspect-ratios and fin widths, the entire two-phase database is compiled in the final Fig.~\ref{twoPhase_TqNu}. Base and wall temperatures, Fig.~\ref{twoPhase_TqNu}(a) and (b), confirm that lower evaporator temperatures are achieved by smaller channel aspect-ratios and fin widths, as less heat is delievered to the evaporator through the narrower base surface, when the heat flux is maintained constant. Therefore, the present results support the conclusion that, for a given heat flux and evaporator width, accommodating multiple low-aspect-ratio microchannels guarantees lower temperatures than a few or one only high-aspect-ratio microchannel; this aspect is discussed further in Sec.~\ref{Sec:disc}. This result is in agreement with the experiments of \citet{alZaidi2021}, who found that multi-microchannel evaporators with lower aspect-ratio channels yield smaller wall superheat for the same base heat flux. The heat flux at the wall-fluid boundary, Fig.~\ref{twoPhase_TqNu}(c)-(e), increases with the fin width as more heat must be delivered to the fluid through the same wall-fluid contact area. The heat flux through the horizontal wall, Fig.~\ref{twoPhase_TqNu}(d), increases monotonically with $\epsilon$ owing to the increasingly wider evaporating liquid film. A similar trend is observed over the vertical wall as the aspect-ratio is reduced from $\epsilon=4$ to $\epsilon=1$, see Fig.~\ref{twoPhase_TqNu}(e), however a non-monotonic behavior of $\overline{\overline{q_{w,v}}}$ occurs as $\epsilon \rightarrow 0$, because the bubble grows proressively more slowly due to the lower wall temperatures, and thus extended dry vapour patches appear over the vertical wall; see wet area fractions in Fig.~\ref{twoPhase_TqNu}(k). This effect is amplified for thinner evaporator fins, as temperatures are lower thus decreasing further the bubble growth rate. Note that the average two-phase heat flux reaches values as high as $1\,\mathrm{MW/m^2}$, corresponding to above $3\,\mathrm{W}$ dissipated by the fluid. This value is about 10 times larger than the heat load applied to the evaporator base, which is possible because the evaporator is not at steady-state, and liquid evaporation is fed by the sensible heat stored within the evaporator walls. It is worth inspecting the plots of the wet area fraction in Fig.~\ref{twoPhase_TqNu}(i)-(k) before discussing those of the Nusselt number. Over the horizontal wall, dry patches are small because when $\epsilon<1$ surface tension forces leave a thick film, see for example Fig.~\ref{crossSections}(a) and (b), whereas when $\epsilon>1$ the bubble grows faster thus depositing a thicker film that better resists dryout. Over the vertical wall, channels with $\epsilon>1$ leave a thick film, whereas when $\epsilon<1$ large dry patches appear (Fig.~\ref{crossSections}(a)) because the bubble propagates more slowly and the thin film deposited by the bubble nose dewets rapidly.
The wet area fraction trends have a direct impact on the Nusselt number plots in Fig.~\ref{twoPhase_TqNu}(f)-(h). Since dry patches over the horizontal wall are small, $\overline{\overline{\mathrm{Nu_{w,h}}}}$ shows a monotonic ascending trend when increasing $\epsilon$, because an increasingly larger fraction of the channel perimeter is covered by a thin film which promotes heat transfer. For the same reason, an analogous ascending trend occurs for $\overline{\overline{\mathrm{Nu_{w,v}}}}$ when decreasing the aspect-ratio from $\epsilon=4$ to about $\epsilon=0.5$. However, this trend changes when $\epsilon<0.5$ due to a sudden drop of $\overline{\overline{\mathrm{Nu_{w,v}}}}$. The latter happens due to the large dry patches that form over the vertical walls when $\epsilon \rightarrow 0$, as explained above. The fall of $\overline{\overline{\mathrm{Nu_{w,v}}}}$ at low aspect-ratios is mitigated by larger evaporator fins, which maintain larger temperatures thus promoting faster bubbles and thicker liquid films. The resulting trends of the average Nusselt number versus $\epsilon$ reported in Fig.~\ref{twoPhase_TqNu}(f) are mixed. For smaller evaporator fins, the overall Nusselt number increases somewhat monotonically with the channel aspect-ratio, due to the steep increase of the wall wet fraction. However, for thicker fins, the drop of $\overline{\overline{\mathrm{Nu_{w,v}}}}$ as $\epsilon \rightarrow 0$ is less severe and a nonmonotonic trend of $\overline{\overline{\mathrm{Nu_{w}}}}$ for increasing aspect-ratios is observed.

\citet{alZaidi2021} reported an increasing trend of the heat transfer coefficient when increasing $\epsilon$ in the range $\epsilon=0.5-2$, using the fluid HFE7100 in $D_h=0.46\,\mathrm{mm}$ channels with $W_f=0.1\,\mathrm{mm}$. They calculated the heat transfer coefficient from temperature and heat flux measurements via thermocouples installed below the horizontal wall of the microchannel, and therefore their heat transfer trends should be compared with our numerical trends for $\overline{\overline{\mathrm{Nu_{w,h}}}}$, Fig.~\ref{twoPhase_TqNu}(g). The trends are in good agreement and the numerical results confirm \citet{alZaidi2021} interpretation that the higher heat transfer coefficient achieved by $\epsilon=2$ can be explained with the presence of an extended evaporating film over the horizontal wall. 

\section{Discussion}
\label{Sec:disc}

The results presented in the previous section emphasise that there are two relevant parameters to describe the performance of an evaporator, the average Nusselt number $\overline{\overline{\mathrm{Nu_{w}}}}$ and the base temperature $\overline{\overline{T_b}}$, which do not necessarily follow the same trends when varying the channel aspect-ratio $\epsilon$. 

The single-phase results provided in Fig.~\ref{shahLondon} show that $\overline{\overline{\mathrm{Nu_{w}}}}$ drops by about 33\% when increasing the aspect-ratio from $\epsilon=0.25$ to 4, to which it corresponds a three/fourfold increase of the base superheat, see Fig.~\ref{singlePhase_TqNu}(a), due to the combined effect of lower $\overline{\overline{\mathrm{Nu_{w}}}}$ and higher fluid temperature. Nonetheless, the single-phase results exhibit a systematic trend of increasing $\overline{\overline{\mathrm{Nu_{w}}}}$ and decreasing $\overline{\overline{T_b}}$ when $\epsilon<1$, thus suggesting that multichannel heat sinks with low aspect-ratio channels ensure better heat transfer and lower base temperatures, when operating in single-phase regime. 

The situation is less clear when the heat sink operates in two-phase flow boiling conditions, because Fig.~\ref{twoPhase_TqNu}(f) indicates that the Nusselt number exhibits non-monotonic trends versus $\epsilon$, and the base temperatures reported in Fig.~\ref{twoPhase_TqNu}(a) are not at steady-state and thus some kind of model is necessary to extrapolate their values to steady-state conditions. If the heat transfer coefficients over the longer vertical walls for $\epsilon<1$ were of similar magnitudes to those over corresponding longer horizontal walls for $\epsilon>1$, it would be expected that $\overline{\overline{\mathrm{Nu_{w}}}}$ increase monotonically for $\epsilon \rightarrow 0$, as observed in the single-phase configuration. However, the two-phase Nusselt numbers reported in Fig.~\ref{twoPhase_TqNu}(f) show mixed trends when varying $\epsilon$ and the fin width $W_f$, as such it is not possible to draw a firm conclusion about which aspect-ratio maximises two-phase heat transfer. In microchannel two-phase flow, where slug flow and annular flow are dominant flow patterns \cite{ong2011}, $\overline{\overline{\mathrm{Nu_{w}}}}$ is directly related to the thickness and distribution of liquid films and dry vapour patches over the heated walls, which depend on a number of hydrodynamic and thermodynamic parameters. Even so, the beneficial effect of a higher Nusselt number for a specific value of $\epsilon$ may be outweighed by the intrinsic thermal resistance of the whole evaporator \cite{alZaidi2021}. 

To further investigate this aspect, we extend here the heat transfer model of the evaporator developed in Sec.~\ref{Sec:model} to account for a two-phase flow, where we assume that $\overline{\overline{T_f}}=T_{sat}$, and consider an evaporator of total base width $W_{tot}$ (base area $A_b=W_{tot} L$), featuring $N_{ch}$ parallel microchannels. Therefore, Eq.~\eqref{Eq:Tbmodel} can be rewritten as: 
\begin{equation}
\overline{\overline{T_b}}=T_{sat}+q_b A_b R_{tot}
\label{Eq:Tbmodeltp}
\end{equation}
where $R_{tot}=R_b +R_w$ and these two thermal resistances are now expressed for the entire evaporator width as:
\begin{equation}
R_b=\frac{H_b}{\lambda_s A_b}, \quad R_w=\frac{R_{w,h} R_{w,v}}{R_{w,h}+ R_{w,v}} \frac{1}{N_{ch}}
\end{equation}
where $N_{ch}=W_{tot}/(W_{ch}+W_f)$, therefore the number of microchannels that can be accommodated depends on their aspect-ratio and the fin width. Equation~\eqref{Eq:Tbmodeltp} can therefore be rewritten as:
\begin{equation}
\overline{\overline{T_b}}=T_{sat}+\frac{q_b H_b} {\lambda_s} + q_b A_b R_w
\label{Eq:Tbmodeltp2}
\end{equation}
where the heat convection resistance at the channel wall, $R_w$, is the only parameter dependent on $\epsilon$. Using the same expressions for the heat convection resistance over the vertical and horizontal walls of the channel developed in Sec.~\ref{Sec:model}, and taking $h_{w,h}=h_{w,v}=h_w$, $R_w$  can be expressed as:
\begin{equation}
R_w=\frac{\epsilon}{A_b} \frac{1+2W_f/[D_h(1+\epsilon)]}{h_w(\epsilon+2\eta)}
\label{Eq:Rw}
\end{equation}
which shows that $R_w \sim \epsilon$ such that the thermal resistance decreases as the aspect-ratio is reduced, and so does the evaporator base temperature by virtue of Eq.~\eqref{Eq:Tbmodeltp}. It is possible to use Eq.~\eqref{Eq:Rw} to estimate the resistance to convective heat transfer per unit area, $R_w A_b$, for the two-phase configuration studied in this paper. Using $W_f=D_h/8$ and $h_w$ from the data in Fig.~\ref{twoPhase_TqNu}(f), $R_w A_b=5.7\,\mathrm{[W/(mm^2\, K)]^{-1}}$ for $\epsilon=0.25$ and $R_w A_b=8.7\,\mathrm{[W/(mm^2\, K)]^{-1}}$ for $\epsilon=4$. Therefore, despite $\overline{\overline{\mathrm{Nu_{w}}}} \approx 27$ for $\epsilon=4$ while $\overline{\overline{\mathrm{Nu_{w}}}} \approx 12$ for $\epsilon=0.25$ (see data for $W_f=D_h/8$ in Fig.~\ref{twoPhase_TqNu}(f)), the thermal resistance to convective heat transfer $R_w$ is still much lower for the lower aspect-ratio channel, and Eq.~\eqref{Eq:Tbmodeltp} suggests that, at steady-state, $\overline{\overline{T_b}}=373.7\,\mathrm{K}$ for $\epsilon=0.25$, versus $\overline{\overline{T_b}}=374\,\mathrm{K}$ for $\epsilon=4$.

In summary, though our two-phase numerical results and steady-state model outlined above seem to suggest that microchannels with $\epsilon <1$ promote lower evaporator temperatures, no absolute answer emerges as to which channel configuration yields the best heat transfer or lowest base temperature. This is expected to depend on the interplay of heat transfer within the solid regions of the evaporator, bubble dynamics and cross-sectional liquid film morphology, which should all be incorporated into a novel three-zone model \cite{3zones1,magnini2017c} to predict boiling heat transfer in noncircular channels, to provide physics-based guidelines for the design of multichannel evaporators.

\section{Conclusions}
\label{Sec:concl}

Flow boiling in a multi-microchannel evaporator was simulated by modelling one single channel and the surrounding walls. The opensource software OpenFOAM v2106 and the built-in geometric Volume Of Fluid method were employed, with self-developed functions improving the estimation of the surface tension and phase-change rate. Square and rectangular microchannels were considered, with aspect-ratios varying in the range $\epsilon=0.25-4$ and widths $W_f=D_h/8-D_h$ of the wall separating adjacent channels. The channel hydraulic diameter, base heat flux, mass flux and fluid properties were maintained constant throughout this work, and water at the saturation temperature of $T_{sat}=100\,\mathrm{^\circ C}$ was the working fluid. The channel walls were set as hydrophilic. The analysis of both single-phase and two-phase fluid dynamics and heat transfer mechanisms led to the following conclusions:
\begin{itemize}
\item Heat is transferred to the fluid through the bottom wall and the two vertical walls at the sides of the channel, thus the heating configuration corresponds to a three-side heated channel. Conjugate heat transfer and channel shape have a profound impact on heat transfer.
\item In the single-phase regime, the three-side configuration leads to increasing Nusselt numbers as smaller aspect-ratio channels are considered, as the vertical wall contributes twice to the average convective heat transfer performance.
\item For the configuration studied, increasing heat loads are delivered to the fluid when the channel aspect-ratio or fin width are increased. In single-phase flow, this corresponds to increasingly higher evaporator temperatures, and therefore the best heat removal performance are achieved when $\epsilon<1$.
\item In the two-phase regime, the bubble quickly becomes elongated under the conditions studied. Local heat transfer rates are the highest, and evaporator temperatures the lowest, in coincidence with liquid-vapour-solid contact lines and thin liquid films.
\item Extended thin liquid films form over the wider microchannel walls, while thicker films are left over shorter walls. The thickness of the film over each wall depends on the bubble speed. At high aspect-ratios, walls are warmer and the bubble grows faster, leaving thicker liquid films that better resist dryout. At low aspect-ratios, bubbles grow more slowly and the thin film over the wider wall dewets leaving extended dry vapour patches which contribute poorly to heat transfer.
\item The Nusselt number of the two-phase flow is directly related to the thickness and morphology of liquid film and dry patches at the microchannel walls. The trends of $\overline{\overline{\mathrm{Nu_{w}}}}$ versus $\epsilon$ are mixed, although for smaller channel fins an ascending trend for increasing aspect-ratios is apparent.
\item Due to the conjugate heat transfer, the heat transfer coefficient and evaporator base temperature exhibit contrasting trends when varying channel aspect-ratio. The present results and a steady-state heat transfer model for the evaporator suggest that configurations with $\epsilon <1$ promote lower evaporator temperatures even when their $\overline{\overline{\mathrm{Nu_{w}}}}$ is below that achieved for $\epsilon >1$, with the higher heat transfer coefficient of the latter being outweighed by the larger overall thermal resistance of the system.
\end{itemize}
This work suggests that it is difficult to draw a general conclusion about which channel aspect-ratio maximises boiling heat transfer, or minimises the evaporator base temperature. This is the result of the interplay among conjugate heat transfer in the solid, bubble, liquid film dynamics and two-phase heat transfer, and only a prediction model incorporating all relevant hydrodynamics and heat exchange processes can provide an optimised heat sink configuration, which is expected to be case-dependent and vary according to operating conditions and working fluid.

\section*{Acknowledgements}

This work is supported by the UK Engineering \& Physical Sciences Research Council (EPSRC), through the BONSAI (EP/T033398/1) grant. Calculations were performed using ARCHER2 UK National Supercomputing Service (archer2.ac.uk), and using the Sulis Tier-2 HPC platform hosted by the Scientific Computing Research Technology Platform at the University of Warwick. Sulis is funded by EPSRC Grant EP/T022108/1 and the HPC Midlands+ consortium. 


\bibliographystyle{model1-num-names}
\bibliography{bibliography}

\begin{thebibliography}{43}
\expandafter\ifx\csname natexlab\endcsname\relax\def\natexlab#1{#1}\fi
\providecommand{\url}[1]{\texttt{#1}}
\providecommand{\href}[2]{#2}
\providecommand{\path}[1]{#1}
\providecommand{\DOIprefix}{doi:}
\providecommand{\ArXivprefix}{arXiv:}
\providecommand{\URLprefix}{URL: }
\providecommand{\Pubmedprefix}{pmid:}
\providecommand{\doi}[1]{\href{http://dx.doi.org/#1}{\path{#1}}}
\providecommand{\Pubmed}[1]{\href{pmid:#1}{\path{#1}}}
\providecommand{\bibinfo}[2]{#2}
\ifx\xfnm\relax \def\xfnm[#1]{\unskip,\space#1}\fi
\bibitem[{Karayiannis and Mahmoud(2017)}]{karayiannis2017}
\bibinfo{author}{T.~G. Karayiannis}, \bibinfo{author}{M.~M. Mahmoud},
\newblock \bibinfo{title}{Flow boiling in microchannels: Fundamentals and
  applications},
\newblock \bibinfo{journal}{Appl. Therm. Eng.} \bibinfo{volume}{115}
  (\bibinfo{year}{2017}) \bibinfo{pages}{1372 -- 1397}.
\bibitem[{Tullius et~al.(2011)Tullius, Vajtai, and Bayazitoglu}]{tullius2011}
\bibinfo{author}{J.~F. Tullius}, \bibinfo{author}{R.~Vajtai},
  \bibinfo{author}{Y.~Bayazitoglu},
\newblock \bibinfo{title}{A review of cooling in microchannels},
\newblock \bibinfo{journal}{Heat Transf. Eng.} \bibinfo{volume}{32}
  (\bibinfo{year}{2011}) \bibinfo{pages}{527--541}.
\bibitem[{Agostini et~al.(2007)Agostini, Fabbri, Park, Wojtan, Thome, and
  Michel}]{agostini2007}
\bibinfo{author}{B.~Agostini}, \bibinfo{author}{M.~Fabbri},
  \bibinfo{author}{J.~E. Park}, \bibinfo{author}{L.~Wojtan},
  \bibinfo{author}{J.~R. Thome}, \bibinfo{author}{B.~Michel},
\newblock \bibinfo{title}{State of the art of high heat flux cooling
  technologies},
\newblock \bibinfo{journal}{Heat Transf. Eng.} \bibinfo{volume}{28}
  (\bibinfo{year}{2007}) \bibinfo{pages}{258--281}.
\bibitem[{Cheng and Xia(2017)}]{cheng2017}
\bibinfo{author}{L.~Cheng}, \bibinfo{author}{G.~Xia},
\newblock \bibinfo{title}{Fundamental issues, mechanisms and models of flow
  boiling heat transfer in microscale channels},
\newblock \bibinfo{journal}{Int. J. Heat Mass Transf.} \bibinfo{volume}{108}
  (\bibinfo{year}{2017}) \bibinfo{pages}{97 -- 127}.
\bibitem[{Al-Zaidi et~al.(2021)Al-Zaidi, Mahmoud, and
  Karayiannis}]{alZaidi2021}
\bibinfo{author}{A.~H. Al-Zaidi}, \bibinfo{author}{M.~M. Mahmoud},
  \bibinfo{author}{T.~G. Karayiannis},
\newblock \bibinfo{title}{Effect of aspect ratio on flow boiling
  characteristics in microchannels},
\newblock \bibinfo{journal}{Int. J. Heat Mass Transf.} \bibinfo{volume}{164}
  (\bibinfo{year}{2021}) \bibinfo{pages}{120587}.
\bibitem[{Harirchian and Garimella(2009)}]{harirchian2009}
\bibinfo{author}{T.~Harirchian}, \bibinfo{author}{S.~V. Garimella},
\newblock \bibinfo{title}{The critical role of channel cross-sectional area in
  microchannel flow boiling heat transfer},
\newblock \bibinfo{journal}{Int. J. Multiph. Flow} \bibinfo{volume}{35}
  (\bibinfo{year}{2009}) \bibinfo{pages}{904 -- 913}.
\bibitem[{Wong et~al.(1995)Wong, Radke, and Morris}]{wong1995a}
\bibinfo{author}{H.~Wong}, \bibinfo{author}{C.~J. Radke},
  \bibinfo{author}{S.~Morris},
\newblock \bibinfo{title}{The motion of long bubbles in polygonal capillaries.
  \mbox{P}art 1. \mbox{T}hin films},
\newblock \bibinfo{journal}{J. Fluid Mech.} \bibinfo{volume}{292}
  (\bibinfo{year}{1995}) \bibinfo{pages}{71--94}.
\bibitem[{de~L\'ozar et~al.(2008)de~L\'ozar, Juel, and Hazel}]{deLozar2008}
\bibinfo{author}{A.~de~L\'ozar}, \bibinfo{author}{A.~Juel},
  \bibinfo{author}{A.~L. Hazel},
\newblock \bibinfo{title}{The steady propagation of an air finger into a
  rectangular tube},
\newblock \bibinfo{journal}{J. Fluid Mech.} \bibinfo{volume}{614}
  (\bibinfo{year}{2008}) \bibinfo{pages}{173--195}.
\bibitem[{Magnini et~al.(2022)Magnini, Municchi, El~Mellas, and
  Icardi}]{magnini2022a}
\bibinfo{author}{M.~Magnini}, \bibinfo{author}{F.~Municchi},
  \bibinfo{author}{I.~El~Mellas}, \bibinfo{author}{M.~Icardi},
\newblock \bibinfo{title}{Liquid film distribution around long gas bubbles
  propagating in rectangular capillaries},
\newblock \bibinfo{journal}{Int. J. Multiph. Flow} \bibinfo{volume}{148}
  (\bibinfo{year}{2022}) \bibinfo{pages}{103939}.
\bibitem[{Han et~al.(2012)Han, Shikazono, and Kasagi}]{han2012}
\bibinfo{author}{Y.~Han}, \bibinfo{author}{N.~Shikazono},
  \bibinfo{author}{N.~Kasagi},
\newblock \bibinfo{title}{The effect of liquid film evaporation on flow boiling
  heat transfer in a microtube},
\newblock \bibinfo{journal}{Int. J. Heat Mass Transf.} \bibinfo{volume}{55}
  (\bibinfo{year}{2012}) \bibinfo{pages}{547--555}.
\bibitem[{Rao and Peles(2015)}]{rao2015}
\bibinfo{author}{S.~R. Rao}, \bibinfo{author}{Y.~Peles},
\newblock \bibinfo{title}{Spatiotemporally resolved heat transfer measurements
  for flow boiling in microchannels},
\newblock \bibinfo{journal}{Int. J. Heat Mass Transf.} \bibinfo{volume}{89}
  (\bibinfo{year}{2015}) \bibinfo{pages}{482--493}.
\bibitem[{Ferrari et~al.(2018)Ferrari, Magnini, and Thome}]{ferrari2018}
\bibinfo{author}{A.~Ferrari}, \bibinfo{author}{M.~Magnini},
  \bibinfo{author}{J.~R. Thome},
\newblock \bibinfo{title}{Numerical analysis of slug flow boiling in square
  microchannels},
\newblock \bibinfo{journal}{Int. J. Heat Mass Transf.} \bibinfo{volume}{123}
  (\bibinfo{year}{2018}) \bibinfo{pages}{928--944}.
\bibitem[{Magnini and Matar(2020)}]{magnini2020a}
\bibinfo{author}{M.~Magnini}, \bibinfo{author}{O.~K. Matar},
\newblock \bibinfo{title}{Numerical study of the impact of the channel shape on
  microchannel boiling heat transfer},
\newblock \bibinfo{journal}{Int. J. Heat Mass Transf.} \bibinfo{volume}{150}
  (\bibinfo{year}{2020}) \bibinfo{pages}{119322}.
\bibitem[{Vontas et~al.(2021)Vontas, Andredaki, Georgoulas, Mich\'e, and
  Marengo}]{vontas2021}
\bibinfo{author}{K.~Vontas}, \bibinfo{author}{M.~Andredaki},
  \bibinfo{author}{A.~Georgoulas}, \bibinfo{author}{N.~Mich\'e},
  \bibinfo{author}{M.~Marengo},
\newblock \bibinfo{title}{The effect of hydraulic diameter on flow boiling
  within single rectangular microchannels and comparison of heat sink
  configuration of a single and multiple microchannels},
\newblock \bibinfo{journal}{Energies} \bibinfo{volume}{14}
  (\bibinfo{year}{2021}) \bibinfo{pages}{6641}.
\bibitem[{Lin et~al.(2021)Lin, Li, Luo, Li, Luo, Kabelac, Cao, and
  Minkowycz}]{lin2021}
\bibinfo{author}{Y.~Lin}, \bibinfo{author}{J.~Li}, \bibinfo{author}{Y.~Luo},
  \bibinfo{author}{W.~Li}, \bibinfo{author}{X.~Luo},
  \bibinfo{author}{S.~Kabelac}, \bibinfo{author}{Y.~Cao},
  \bibinfo{author}{W.~J. Minkowycz},
\newblock \bibinfo{title}{Conjugate heat transfer analysis of bubble growth
  during flow boiling in a rectangular microchannel},
\newblock \bibinfo{journal}{Int. J. Heat Mass Transf.} \bibinfo{volume}{181}
  (\bibinfo{year}{2021}) \bibinfo{pages}{121828}.
\bibitem[{Szczukiewicz et~al.(2014)Szczukiewicz, Magnini, and
  Thome}]{szczukiewicz2014}
\bibinfo{author}{S.~Szczukiewicz}, \bibinfo{author}{M.~Magnini},
  \bibinfo{author}{J.~R. Thome},
\newblock \bibinfo{title}{Proposed models, ongoing experiments, and latest
  numerical simulations of microchannel two-phase flow boiling},
\newblock \bibinfo{journal}{Int. J. Multiph. Flow} \bibinfo{volume}{59}
  (\bibinfo{year}{2014}) \bibinfo{pages}{84--101}.
\bibitem[{Roenby et~al.(2016)Roenby, Bredmose, and Jasak}]{roenby2016}
\bibinfo{author}{J.~Roenby}, \bibinfo{author}{H.~Bredmose},
  \bibinfo{author}{H.~Jasak},
\newblock \bibinfo{title}{A computational method for sharp interface
  advection},
\newblock \bibinfo{journal}{R. Soc. Open Sci.} \bibinfo{volume}{3}
  (\bibinfo{year}{2016}) \bibinfo{pages}{160405}.
\bibitem[{Tryggvason et~al.(2011)Tryggvason, Scardovelli, and
  Zaleski}]{scardo_book}
\bibinfo{author}{G.~Tryggvason}, \bibinfo{author}{R.~Scardovelli},
  \bibinfo{author}{S.~Zaleski}, \bibinfo{title}{Direct numerical simulations of
  gas-liquid multiphase flows}, \bibinfo{publisher}{Cambridge University
  Press}, \bibinfo{address}{Cambridge}, \bibinfo{year}{2011}.
\bibitem[{Brackbill et~al.(1992)Brackbill, Kothe, and Zemach}]{brackbill}
\bibinfo{author}{J.~U. Brackbill}, \bibinfo{author}{D.~B. Kothe},
  \bibinfo{author}{C.~Zemach},
\newblock \bibinfo{title}{A continuum method for modeling surface tension},
\newblock \bibinfo{journal}{J. Comput. Phys.} \bibinfo{volume}{100}
  (\bibinfo{year}{1992}) \bibinfo{pages}{335--354}.
\bibitem[{Hoang et~al.(2013)Hoang, van Steijn, Portela, Kreutzer, and
  Kleijn}]{hoang2013}
\bibinfo{author}{D.~A. Hoang}, \bibinfo{author}{V.~van Steijn},
  \bibinfo{author}{L.~M. Portela}, \bibinfo{author}{M.~T. Kreutzer},
  \bibinfo{author}{C.~R. Kleijn},
\newblock \bibinfo{title}{Benchmark numerical simulations of segmented
  two-phase flows in microchannels using the \mbox{V}olume of \mbox{F}luid
  method},
\newblock \bibinfo{journal}{Comput. Fluids} \bibinfo{volume}{86}
  (\bibinfo{year}{2013}) \bibinfo{pages}{28--36}.
\bibitem[{Hardt and Wondra(2008)}]{hardt}
\bibinfo{author}{S.~Hardt}, \bibinfo{author}{F.~Wondra},
\newblock \bibinfo{title}{Evaporation model for interfacial flows based on a
  continuum-field representation of the source terms},
\newblock \bibinfo{journal}{J. Comput. Phys.} \bibinfo{volume}{227}
  (\bibinfo{year}{2008}) \bibinfo{pages}{5871--5895}.
\bibitem[{Carey(1992)}]{carey}
\bibinfo{author}{V.~P. Carey}, \bibinfo{title}{Liquid-vapor phase change
  phenomena}, \bibinfo{publisher}{Taylor and Francis}, \bibinfo{year}{1992}.
\bibitem[{Tanasawa(1991)}]{tanasawa}
\bibinfo{author}{I.~Tanasawa},
\newblock \bibinfo{title}{Advances in condensation heat transfer},
\newblock in: \bibinfo{editor}{J.~P. Hartnett}, \bibinfo{editor}{T.~F. Irvine}
  (Eds.), \bibinfo{booktitle}{Advances in Heat Transfer},
  \bibinfo{publisher}{Academic Press}, \bibinfo{address}{San Diego},
  \bibinfo{year}{1991}.
\bibitem[{Magnini et~al.(2013)Magnini, Pulvirenti, and Thome}]{magnini2013a}
\bibinfo{author}{M.~Magnini}, \bibinfo{author}{B.~Pulvirenti},
  \bibinfo{author}{J.~R. Thome},
\newblock \bibinfo{title}{Numerical investigation of hydrodynamics and heat
  transfer of elongated bubbles during flow boiling in a microchannel},
\newblock \bibinfo{journal}{Int. J. Heat Mass Transf.} \bibinfo{volume}{59}
  (\bibinfo{year}{2013}) \bibinfo{pages}{451--471}.
\bibitem[{Weller(2008)}]{weller2008}
\bibinfo{author}{H.~G. Weller},
\newblock \bibinfo{title}{A new approach to \mbox{VOF}-based interface
  capturing methods for incompressible and compressible flows},
\newblock \bibinfo{journal}{OpenCFD Ltd. Report TR/HGW/04}
  (\bibinfo{year}{2008}).
\bibitem[{Deshpande et~al.(2012)Deshpande, Anumolu, and
  Trujillo}]{deshpande2012}
\bibinfo{author}{S.~S. Deshpande}, \bibinfo{author}{L.~Anumolu},
  \bibinfo{author}{M.~F. Trujillo},
\newblock \bibinfo{title}{Evaluating the performance of the two-phase flow
  solver inter\mbox{F}oam},
\newblock \bibinfo{journal}{Comput. Sci. Discov.} \bibinfo{volume}{5}
  (\bibinfo{year}{2012}) \bibinfo{pages}{1--36}.
\bibitem[{Scheufler and Roenby(2019)}]{scheufler2019}
\bibinfo{author}{H.~Scheufler}, \bibinfo{author}{J.~Roenby},
\newblock \bibinfo{title}{Accurate and efficient surface reconstruction from
  volume fraction data on general meshes},
\newblock \bibinfo{journal}{J. Comput. Phys.} \bibinfo{volume}{383}
  (\bibinfo{year}{2019}) \bibinfo{pages}{1--23}.
\bibitem[{van Leer(1979)}]{muscl}
\bibinfo{author}{B.~van Leer},
\newblock \bibinfo{title}{Towards the ultimate conservative difference scheme.
  \mbox{V}. \mbox{A} second-order sequel to \mbox{G}odunov's method},
\newblock \bibinfo{journal}{J. Comput. Phys.} \bibinfo{volume}{32}
  (\bibinfo{year}{1979}) \bibinfo{pages}{101--136}.
\bibitem[{Issa(1985)}]{piso}
\bibinfo{author}{R.~I. Issa},
\newblock \bibinfo{title}{Solution of the implicitly discretized fluid flow
  equations by operator-splitting},
\newblock \bibinfo{journal}{J. Comput. Phys.} \bibinfo{volume}{62}
  (\bibinfo{year}{1985}) \bibinfo{pages}{40--65}.
\bibitem[{Scriven(1959)}]{scriven}
\bibinfo{author}{L.~E. Scriven},
\newblock \bibinfo{title}{On the dynamics of phase growth},
\newblock \bibinfo{journal}{Chem. Eng. Sci.} \bibinfo{volume}{10}
  (\bibinfo{year}{1959}) \bibinfo{pages}{1--13}.
\bibitem[{Mukherjee et~al.(2011)Mukherjee, Kandlikar, and Edel}]{mukherjee2011}
\bibinfo{author}{A.~Mukherjee}, \bibinfo{author}{S.~G. Kandlikar},
  \bibinfo{author}{Z.~J. Edel},
\newblock \bibinfo{title}{Numerical study of bubble growth and wall heat
  transfer during flow boiling in a microchannel},
\newblock \bibinfo{journal}{Int. J. Heat Mass Transf.} \bibinfo{volume}{54}
  (\bibinfo{year}{2011}) \bibinfo{pages}{3702--3718}.
\bibitem[{Gamet et~al.(2020)Gamet, Scala, Roenby, Scheufler, and
  Pierson}]{gamet2020}
\bibinfo{author}{L.~Gamet}, \bibinfo{author}{M.~Scala},
  \bibinfo{author}{J.~Roenby}, \bibinfo{author}{H.~Scheufler},
  \bibinfo{author}{J.-L. Pierson},
\newblock \bibinfo{title}{Validation of volume-of-fluid \mbox{OpenFOAM}
  iso\mbox{A}dvector solvers using single bubble benchmarks},
\newblock \bibinfo{journal}{Comput. Fluids} \bibinfo{volume}{213}
  (\bibinfo{year}{2020}) \bibinfo{pages}{104722}.
\bibitem[{Abadie et~al.(2015)Abadie, Aubin, and Legendre}]{abadie2015}
\bibinfo{author}{T.~Abadie}, \bibinfo{author}{J.~Aubin},
  \bibinfo{author}{D.~Legendre},
\newblock \bibinfo{title}{On the combined effects of surface tension force
  calculation and interface advection on spurious currents within \mbox{V}olume
  of \mbox{F}luid and \mbox{L}evel \mbox{S}et frameworks},
\newblock \bibinfo{journal}{J. Comput. Phys.} \bibinfo{volume}{297}
  (\bibinfo{year}{2015}) \bibinfo{pages}{611--636}.
\bibitem[{Falsetti et~al.(2018)Falsetti, Magnini, and Thome}]{falsetti2018}
\bibinfo{author}{C.~Falsetti}, \bibinfo{author}{M.~Magnini},
  \bibinfo{author}{J.~R. Thome},
\newblock \bibinfo{title}{Hydrodynamic and thermal analysis of a micro-pin fin
  evaporator for on-chip two-phase cooling of high density power
  micro-electronics},
\newblock \bibinfo{journal}{Appl. Therm. Eng.} \bibinfo{volume}{130}
  (\bibinfo{year}{2018}) \bibinfo{pages}{1425--1439}.
\bibitem[{Shah and London(1971)}]{shahLondon}
\bibinfo{author}{R.~K. Shah}, \bibinfo{author}{A.~L. London},
  \bibinfo{title}{Laminar Flow Forced Convection Heat Transfer and Flow
  Friction in Straight and Curved Ducts. \mbox{A} Summary of Analytical
  Solutions}, \bibinfo{type}{Technical Report} \bibinfo{number}{AD736260},
  Stanford University, \bibinfo{year}{1971}.
\bibitem[{Bretherton(1961)}]{bretherton1961}
\bibinfo{author}{F.~P. Bretherton},
\newblock \bibinfo{title}{The motion of long bubbles in tubes},
\newblock \bibinfo{journal}{J. Fluid Mech.} \bibinfo{volume}{10}
  (\bibinfo{year}{1961}) \bibinfo{pages}{166--188}.
\bibitem[{Magnini and Thome(2016)}]{magnini2016b}
\bibinfo{author}{M.~Magnini}, \bibinfo{author}{J.~R. Thome},
\newblock \bibinfo{title}{A \mbox{CFD} study of the parameters influencing heat
  transfer in microchannel slug flow boiling},
\newblock \bibinfo{journal}{Int. J. Therm. Sci.} \bibinfo{volume}{110}
  (\bibinfo{year}{2016}) \bibinfo{pages}{119--136}.
\bibitem[{Hazel and Heil(2002)}]{hazel2002}
\bibinfo{author}{A.~L. Hazel}, \bibinfo{author}{M.~Heil},
\newblock \bibinfo{title}{The steady propagation of a semi-infinite bubble into
  a tube of elliptical or rectangular cross-section},
\newblock \bibinfo{journal}{J. Fluid Mech.} \bibinfo{volume}{470}
  (\bibinfo{year}{2002}) \bibinfo{pages}{91--114}.
\bibitem[{Khodaparast et~al.(2017)Khodaparast, Kim, Silpe, and
  Stone}]{khodaparast2017}
\bibinfo{author}{S.~Khodaparast}, \bibinfo{author}{M.~K. Kim},
  \bibinfo{author}{J.~Silpe}, \bibinfo{author}{H.~A. Stone},
\newblock \bibinfo{title}{Bubble-driven detachment of bacteria from confined
  micro-geometries},
\newblock \bibinfo{journal}{Environ. Sci. Technol.} \bibinfo{volume}{51}
  (\bibinfo{year}{2017}) \bibinfo{pages}{1340--1347}.
\bibitem[{Khodaparast et~al.(2018)Khodaparast, Atasi, Deblais, Scheid, and
  Stone}]{khodaparast2018}
\bibinfo{author}{S.~Khodaparast}, \bibinfo{author}{O.~Atasi},
  \bibinfo{author}{A.~Deblais}, \bibinfo{author}{B.~Scheid},
  \bibinfo{author}{H.~A. Stone},
\newblock \bibinfo{title}{Dewetting of thin liquid films surrounding long
  bubbles in microchannels},
\newblock \bibinfo{journal}{Langmuir} \bibinfo{volume}{34}
  (\bibinfo{year}{2018}) \bibinfo{pages}{1363--1370}.
\bibitem[{Ong and Thome(2011)}]{ong2011}
\bibinfo{author}{C.~L. Ong}, \bibinfo{author}{J.~R. Thome},
\newblock \bibinfo{title}{Macro-to-microchannel transition in two-phase flow:
  \mbox{P}art 1 - two-phase flow patterns and film thickness measurements},
\newblock \bibinfo{journal}{Exp. Therm. Fluid Sci.} \bibinfo{volume}{35}
  (\bibinfo{year}{2011}) \bibinfo{pages}{37--47}.
\bibitem[{Thome et~al.(2004)Thome, Dupont, and Jabobi}]{3zones1}
\bibinfo{author}{J.~R. Thome}, \bibinfo{author}{V.~Dupont},
  \bibinfo{author}{A.~M. Jabobi},
\newblock \bibinfo{title}{Heat transfer model for evaporation in microchannels.
  \mbox{P}art \mbox{I}: \mbox{P}resentation of the model},
\newblock \bibinfo{journal}{Int. J. Heat Mass Transf.} \bibinfo{volume}{47}
  (\bibinfo{year}{2004}) \bibinfo{pages}{3375--3385}.
\bibitem[{Magnini and Thome(2017)}]{magnini2017c}
\bibinfo{author}{M.~Magnini}, \bibinfo{author}{J.~R. Thome},
\newblock \bibinfo{title}{An updated three-zone heat transfer model for slug
  flow boiling in microchannels},
\newblock \bibinfo{journal}{Int. J. Multiph. Flow} \bibinfo{volume}{91}
  (\bibinfo{year}{2017}) \bibinfo{pages}{296--314}.

\end{thebibliography}







\end{document}